# Academic vs. biological age in research on academic careers: a large-scale study with implications for scientifically developing systems

**Marek Kwiek**[1] · **Wojciech Roszka**[2]

## Abstract

Biological age is an important sociodemographic factor in studies on academic careers (research productivity, scholarly impact, and collaboration patterns). It is assumed that the academic age, or the time elapsed from the first publication, is a good proxy for biological age. In this study, we analyze the limitations of the proxy in academic career studies, using as an example the entire population of Polish academic scientists and scholars visible in the last decade in global science and holding at least a PhD ($N = 20,569$). The proxy works well for science, technology, engineering, mathematics, and medicine (STEMM) disciplines; however, for non-STEMM disciplines (particularly for humanities and social sciences), it has a dramatically worse performance. This negative conclusion is particularly important for systems that have only recently visible in global academic journals. The micro-level data suggest a delayed participation of social scientists and humanists in global science networks, with practical implications for predicting biological age from academic age. We calculate correlation coefficients, present contingency analysis of academic career stages with academic positions and age groups, and create a linear multivariate regression model. Our research suggests that in scientifically developing countries, academic age as a proxy for biological age should be used more cautiously than in advanced countries: ideally, it should be used only for STEMM disciplines.

---

✉ Marek Kwiek
   kwiekm@amu.edu.pl

   Wojciech Roszka
   wojciech.roszka@ue.poznan.pl

1  Institute for Advanced Studies in Social Sciences and Humanities, UNESCO Chair in Institutional Research and Higher Education Policy, Adam Mickiewicz University of Poznan, Poznan, Poland

2  Poznan University of Economics and Business, Poznan, Poland

🙋 Springer

## Introduction

In the previous decade, research on academic careers conducted at the micro-level of the individual scientist, rather than at the aggregate country level, has increasingly relied on the category of academic age (Abramo et al., 2016; Aksnes et al., 2011a; Gingras et al., 2008). The reason for this is simple: the data on biological age is generally unavailable for large-scale studies. Age is one of the most important variables related to numerous dimensions of academic careers, such as research productivity, international mobility, or international research collaboration (Levin & Stephan 1989, 1991). Therefore, it has often been assumed, somewhat out of necessity, that the academic age (or the time elapsed from the first publication) is a good proxy for the biological age (Robinson-Garcia et al., 2020; Milojević, 2012; Nane et al., 2017). The date of first publication in Scopus or Web of Science can be calculated and used for research purposes for all publishing scientists and scholars at the level of institutions or their clusters, cities, regions, disciplines, journals, and countries.

Access to data on the date of first publication on a large scale (such as, for example, for a whole country) requires rather complex preparation, particularly including a list of individual identifiers of all authors, but it is technically possible. On the other hand, access to data on the biological age of researchers, given the current level of development of large administrative databases encompassing all researchers within a country and access to them for scientific purposes, in practice does not exist (with a few exceptions, including Italy and Norway) (see Rørstad & Aksnes, 2015; Rørstad et al., 2021; Abramo et al., 2011). These also include Poland and a database we have created ("Polish Science Observatory") based on the integration of an administrative register of all scientists and scholars working in public science sectors (approximately 100,000 scientists)—maintained by a national institution created for this purpose (OPI-PIB)—with the database of all Polish publications indexed in Scopus in the decade of 2009–2018 (approximately 380,000). Our Observatory database has been used several times in our extensive studies of academic faculty (e.g.: Kwiek & Roszka 2021a, b, 2022 ). The integration of both databases was performed using probabilistic and deterministic methods and we have described its construction in detail elsewhere (see Kwiek & Roszka, 2021a: 4–6).

In this paper, our goal is to analyze the limitations of using academic age as a proxy for biological age in the example of the entire national science system, for which we consider both biological age and academic age for all researchers affiliated to Polish institutions of higher education, holding at least a PhD degree, and participating in global academic science via global publishing. An approximation of functioning in global science is having at least one publication indexed in the Scopus database in the decade 2009–2018.

Since we have an unambiguously defined biological age from the national registry of scientists for each scientist and we are able to determine the academic age for all of them, we can systematically show the differences between the two types of age depending on the selected key parameters. In other words, in this study, we determine the academic age for each Polish scientist based on the date of their first publication and systemically analyze the differences between academic age and their actual biological age. In addition, we also show how the individually determined academic age differs in practice from the strictly determined biological age depending on independent variables such as gender, scientific discipline, academic position, and institutional type.

Thus, using comprehensive data from an entire national system comprising 100,000 researchers and their 380,000 publications as an example, we estimate the scope of limitations of the use of academic age as a proxy for biological age depending on the selected independent variables and analyze both the practical and methodological implications of using academic age in academic career research.

As our study reveals, these limitations turn out to be greatest for the humanities and social sciences: for these disciplines, the level of correlation of both types of age turn out to be low. Therefore, in the analyzed Polish case, it is not possible to practically derive biological age from academic age in case of scholars publishing in these disciplines, as the correlation coefficient ranges from 0.35 for arts and humanities and 0.43 for economics, econometrics, and finance to 0.49 for psychology. In contrast, for science disciplines, the correlation (and determination) is dramatically higher, and using academic age as a proxy for biological age reduces the risk (correlation coefficient ranges from 0.82 for biochemistry, genetics, and molecular biology to 0.89 for chemistry and 0.90 for immunology and microbiology). Furthermore, correlations and determinations take different values for different individual and institutional parameters, which we explore in greater detail below.

Importantly, our study was conducted for a system of science that is in the pursuit of massive global science participation and has only relatively recently adopted global publishing patterns. Poland's contribution to global science is significant, particularly given the relatively low investment in academic science and the inconsistent and outdated system of promotion in science. Systems of promotion strongly affect publishing patterns (Stephan, 2012) and in Poland it is based on a complicated ladder of successive academic degrees, from doctorate to habilitation to full professorship, accompanied by a parallel, complicated system of academic positions (Kwiek & Szadkowski, 2019). Poland ranks seventeenth in the Scopus database in terms of the number of scientific articles in 2020 (43,618 articles), compared to 12,868 articles in 2000 and 22,362 in 2010 (SciVal 2021). The presence of Polish science in the global system of science is steadily increasing in terms of numbers and percentages, but it is strongly diversified from a disciplinary perspective. While the global visibility of STEMM sciences—particularly chemistry and physics—is stable and solid, the presence of humanities, social sciences and economics in global datasets of indexed journals is relatively small and has been observed only recently. In particular, the number of publications in highly prestigious journals (top 10% and top 1% of journals indexed in Scopus by its CiteScore measure) has been growing more strongly in relatively recent times, which is certainly the effect of a full decade of reforms of science and higher education systems (Antonowicz et al., 2020; Bieliński & Tomczyńska, 2018; Feldy & Kowalczyk, 2020; Kulczycki et al., 2017; Kwiek, 2018a, b; Shaw, 2019).

One of the limitations of our study is the absence of some academic researchers in the Scopus database—the lack of indexed publications attributed to them is related to the delayed entry of the Polish academic system into the global scientific circulation. However, this feature is common to many countries that are scientifically catching-up, in this sense, the empirical part of the paper is relevant to a large part of the world undergoing similar changes in publication patterns over time.

In this paper, academic age is defined as the time that has elapsed between the date of the first publication indexed in the Scopus database (all publication types, any role played in publication) and 2017, our reference year. It can be assumed that in the sciences, the first publication in the database most often appears during the last years of writing a PhD dissertation—that is, in the Polish case, at approximately 30 years of age. Thus, in a large simplification, an academic age of 0 may indicate 30 years of age and academic age of 40 may indicate 70 years of age. In this sense, in a very approximate sense, an academic career would encompass a maximum of 40 years of life: academic age range 0–40 years and biological age range 30–70 years. Such a view of an academic career corresponds to the Polish model, in which there is an extremely strong trend of academic inbreeding—that is, working at the same university from the time of obtaining a doctoral degree, and often graduation; remaining in the system of science for several decades; and a marginal level of admission to academic work at a later age, including new admissions, as part of the transition from the economic or public administration sector to the academic sector. The career model includes entering the workforce at around age of 30 years and working in the sector or leaving academia with no option to return. However, most importantly a long career in the academic sector does not necessarily imply a long period of publishing articles indexed in global bibliometric databases (particularly in non-STEMM fields), which distinguishes the Polish system from systems that have been present longer and more intensively in the global circuit of indexed science.

In the first part of the Results section, we present a detailed estimation of the scale of overlapping between the two types of age in different cross-sections and propose broader generalizations. In the second part of the Results section, using a linear multivariate regression model, we estimate biological age on the basis of academic age and other selected parameters directly (institutional affiliation) or indirectly available in the Scopus database (such as the ASJC discipline dominating in a scientist's 10-year publication portfolio; gender defined by first and subsequent names and possibly by external databases and special gender-ascribing algorithms). Further, we explore the utility of the model by inquiring the extent to which we can simulate biological age based on publication metadata available in global bibliometric databases.

## Key literature

### Biological age in academic career research

Biological age has been an important sociodemographic factor in sociological and bibliometric studies of academic careers (especially of research productivity, scholarly impact, and collaboration patterns) for more than half a century (Cole, 1979; Kyvik, 1990; Kyvik & Olsen, 2008; Lehman, 1953; Levin & Stephan, 1991; Pelz & Andrews, 1976; Stephan & Levin, 1992; Stern, 1978; Zuckerman & Merton, 1973). Specifically, biological age has been studied in the USA, Canada, Norway, Spain, Belgium, Poland, and Italy. Kyvik examined a sample of Norwegian scientists and has shown that productivity reaches a peak when scientists are 45–49, followed by productivity decline (1990: 38), with substantial differences between academic fields. Six different hypotheses were suggested (Kyvik & Olsen, 2008: 441–442) to explain reduced research performance of universities with an aging academic staff, related to either a decline in productivity ("the utility maximizing hypothesis", "the seniority burden hypothesis", and "the cumulative disadvantage hypothesis") or to a decline in creativity of research ("the age decrement hypothesis", "the obsolescence hypothesis", and "the intellectual deadlock hypothesis"). Kyvik and Olsen used three survey data sets on Norwegian professors divided into age cohorts from three decades (1982, 1992, and 2001, with a total sample of 5,367 scientists) and show that the decline in productivity with increasing age may be due to generation effects rather than aging effects.

Further, some cohorts may be more research productive than others due to different competition levels in hiring in their early years and also due to different research funding opportunities at that time. Scientists and scholars hired under different conditions may stay on in academia for decades. As Stephan and Levin (1992: 117) noted in a US context, "the relative attractiveness of careers in science changes over time," bringing young people with different talents, motivations, and abilities to academia (Levin & Stephan, 1991: 126). Academic cohorts may be more or less productive from the moment they have entered the academic profession and some cohorts may have always been characterized by low productivity (Kyvik, 1990: 51). Thus, the academic careers of scientists and scholars are affected by events that occur at the time of obtaining a PhD: in other words, "cohort matters" (Stephan, 2012: 175), with "opportunity structures" shifting over time (Stephan & Levin, 1992: 117).

A large-scale study of 11,500 Norwegian scientists suggested an inverted U-shaped publication pattern: productivity increases with age, reaching a peak late in the career and declining thereafter, with scientists over 60 being significantly less cited than their younger colleagues (Aksnes et al., 2011a: 42–44). In the Norwegian case, however, academic positions have been found to be more important for publication rates than age and gender (Rørstad & Aksnes, 2015: 327), with "professors being by far the most prolific persons" (Aksnes et al., 2011a: 43). Moreover, a study of 6,388 university professors in Quebec (Gingras et al., 2008: 6) indicated that there are two important turning points: productivity first increases sharply with age and then increases at a slower pace at about 40 years of age; and then productivity reaches its peak at about 50 years of age. Another Quebec-focused study with age data reveals (Larivière et al., 2011: 483) that after scientists have crossed the age of 38 years, "women receive, on average, less funding for research than men, are generally less productive in terms of publications, and are at a slight disadvantage in terms of the scientific impact".

In a study of Italian full professors, Abramo et al., (2016: 318) concluded that "as age increases there is a high decline in full professors' productivity". However, professors appointed at a young age were more likely to maintain and increase their productivity than colleagues promoted at a later age (Abramo et al., 2016: 318). A "negative monotonic relationship" between age and research performance was found to accompany a "positive relationship between seniority in rank and performance" (Abramo et al., 2016: 301). In their bibliometric study of Spanish National Research Council scientists, Costas et al., (2010a, 2010b) concluded that the productivity of top- and medium-performing scientists increases or remains stable with age and the productivity of low-performing researchers tends to decrease with age (Costas et al., 2010a, 2010b: 1578). Guns et al. (2019) examined different publication patterns of successive age cohorts at a social sciences and humanities department in Flanders, Belgium, and showed that the oldest cohort appears to maintain the traditional publication pattern focused on book publications.

Biological age was also used in research collaboration studies: there are only few studies of the impact of age on collaboration because few data sets combine biographical and publication, sometimes combined with citation data at the individual level. Collaboration, with data on age, can be studied at the meso-level of institutions; however, studies at the macro-level of countries are rare because of limited data availability (examples include Abramo et al. (2011) and Abramo et al. (2016) on Italian scientists; Norwegian studies based on a comprehensive national database Current Research Information System in Norway (CRISTIN), such as Aksnes et al. (2011b) and Rørstad and Aksnes (2015) and Rørstad et al., 2021; and survey-based studies on Polish (Kwiek, 2015b, 2020) and European (Kwiek, 2018b, 2019) scientists, specifically "internationalists" and "locals" in research by age groups).

Biological age has also figured prominently in studies focused on bibliometric indicators at the micro-level of individual scientists and teams (Costas & Bordons, 2005, 2007; Costas et al., 2010a, 2010b; Sugimoto et al., 2016). For example, Costas and Bordons (2010) proposed a bibliometric classificatory approach to research performance and, referring to biological age, presented evidence that top researchers are younger than the other two research productivity classes (low and medium classes). Finally, Wang and Barabàsi (2021: 39–50) examined the question of when scientists do their greatest work and what determines the timing of peaks, thereby revealing that the peak age of Nobel laureates has increased over time by approximately six years. Biological age and the micro-level have also been applied in our own studies in which we used specifically constructed "individual publication portfolios" based on bibliometric metadata from a decade of publications indexed in Scopus and studied by academic age groups (international collaboration: Kwiek & Roszka, 2021a; man-woman collaboration: Kwiek & Roszka, 2021b; and solo research: Kwiek & Roszka, 2022).

## Academic age as a proxy of biological age thus far

Although biological age is an important independent sociodemographic variable in examinations of academic careers, academic age as its proxy has not been used often in research studies thus far. The major reason for this has been a limited access to first publication data. Except for a relatively limited number of studies briefly summarized above where biological age was actually available, biological age has been studied through two major proxies: (1) academic age related to the date of first publication and (2) academic age related to the date of receiving a PhD.

First, in studies using academic age, the date of the first publication was applied approximately a dozen times (see, e.g., Lee & Bozeman, 2005; Milojević, 2012; Radicchi & Castellano, 2013; Nane et al., 2017; Robinson-Garcia et al., 2020; Liao, 2017; Costas et al., 2015; Aref et al., 2019; Chan & Torgler, 2020; Simoes & Crespo, 2020; Petersen, 2015; Wildgaard, 2015). The samples in these studies were of different sizes, with resultant problems of generalizations from small samples to large national populations in several cases. The small samples used in these studies ranged from 137 scholars in information systems (Liao, 2017) to 472 top economists (Simoes & Crespo, 2020), 473 collaboration profiles of scientists in biology and physics (Petersen, 2015), 512 European researchers from 4 disciplines (Wildgaard, 2015), to 3574 (Nane et al., 2017) and 3596 (Costas et al., 2015) scientists from Quebec. Larger samples included 21,562 scientists from 10 core journals in 5 disciplines (Milojević, 2012); 35,136 scientists with profiles in Google Scholar Citations database (Radicchi & Castellano, 2013); 94,000 scientists across 43 countries (Chan & Torgler, 2020); and as many as 222,925 distinct authors with at least 5 publications (Robinson-Garcia et al., 2020) and 1.7 million authorship records in Web of Science (Aref et al., 2019). However, academic age was used for different purposes and occasionally played only a marginal role in the above research.

Second, the date of receiving a PhD as a means to determine academic age, usually derived from CVs, was used even less intensively and referred to generally smaller samples. Badar et al. (2014) studied 239 faculty members in Pakistan; Perianes-Rodriguez (2014) examined 2,530 economists working in the 81 top world economics departments; Sugimoto et al. (2016) studied 1002 scientists working in top 10 US programs in sociology, political science, and economics; van den Besselaar and Sandström (2016) examined 243 early-career researchers in economics, behavior and education, and psychology who applied to an early career grant program in the Netherlands; and, finally, Coomes et al. (2013) studied 369 scientists from 17 US and Canadian geography departments. Finally, Savage and Olejniczak (2021) conducted a large-scale study of 167,299 faculty members from Ph.D.-granting universities in the United States, using the Academic Analytics commercial database.

Third, and in more detail, academic age was used in various data contexts to support different streams of research:

(1) Milojević (2012) examined the citing behavior of scientists in relation to age, productivity, and collaboration and was interested in cohorts of scientists who are at the same stage of their academic careers and, thus, have the same academic age. She argued that the operationalization of the academic age through publications is a good one because it "represents the length of an author's active engagement within a scientific community" (2012: 3).

(2) Radicchi and Castellano (2013) analyzed bibliometric indicators at the level of individual scientists, explored their population in terms of academic age, and calculated the *h*-index and other parameters of academic publishing using academic age.

(3) Among the 17 author-level bibliometric indicators, Wildgaard (2015: 5) distinguishes between three categories—publication-based indicators, citation-based indicators, and hybrid indicators—and locates academic age in publication-based ones. She defines it simply as the "number of years since first publication by the researcher recorded in the database."

(4) Aref et al. (2019) studied highly mobile researchers or "super-movers", contributing to the "brain circulation" (rather than brain gain/drain) literature and offering, for the first time, a snapshot of their key features by academic age. They show the most common countries of academic destination for major age brackets of super-movers (early career, intermediate, and senior super-movers) or mobile scientists who have had main affiliations in at least three distinct countries. No limitations regarding the use of academic age are discussed, as opposed to other types of limitations such as the limitations of bibliometric data in general (Aref et al., 2019: 59).

(5) Simoes and Crespo (2020: 336) applied academic age to their model of measuring author-level research productivity, assuming that any author-level performance evaluation requires the comparison of scientists in different stages of their careers. Measuring the accumulated performance, favoring authors with longer careers, was contrasted with analyzing performance data per unit of time.

(6) Chan and Torgler (2020) used academic age in the context of gender differences in performance of science elites (top cited scientists) across 43 countries and in 21 academic fields; male top-cited scientists were found to be on average 3.2 years older than female top-cited scientists.

(7) Robinson-Garcia et al. (2020) used academic age in their recent research in task specialization across research careers; they used a data set of over 70,000 publications from PLoS journals (with approximately 350,000 distinct authors) to predict the contributions of over 220,000 authors in over 6 million publications. Their major variable was the four academic career stages. The four career archetypes (junior, early-career, mid-career, and late-career) were used to profile scientists and to examine career trajectories, productivity, and citation impact. The authors estimated researchers' age based on the year of first publication and built the four career stages based on such year (Robinson-Garcia et al., 2020: 18).

Finally, and most importantly for our line of research, Nane et al. (2017) analyzed the prediction of the age of researchers using bibliometric data and focused on the predictive power of various regression models. They assessed how reliable the estimation of the real ages of scholars was, defined as their biological age, based on models that exclusively rely on the various bibliometric indicators in the empirical context of researchers from Quebec for whom biological age was available. The authors believe that their sample ($N = 3574$) is representative for researchers in general (Nane et al. 2017: 714); moreover, the sample represents core global science producing systems in the same manner in which we believe that our sample represents (or is useful to understand) catching-up systems in science. While Canada is classified among "advanced countries" in a ranking of scientific capacity and infrastructure of 76 countries, Poland is classified among "developing countries" (Wagner, 2008: 88). Thus, their paper was a test of the accuracy and validity of the first year of publication as a proxy for biological age. From among numerous bibliometric indicators, the year of the first publication was found to be the best single linear estimator of the ages of individual researchers (Nane et al. 2017: 726). However, it was found to function particularly well when working with large sets of scholars (from an "averaged" viewpoint). Furthermore, the predictive power at the individual observational case was found to be "relatively limited, especially in some fields" (Nane et al. 2017: 726).

The abovementioned paper comes closest to our research in terms of its aims, goals, and approach. However, in our research, we have worked with a much larger data set with administrative, biographical, and publication and citation data ("Polish Science Observatory," with national registry data and Scopus data combined) in which biological age was provided for over 20,000 scientists and scholars. Here, we use STEMM and non-STEMM fields, 24 academic disciplines, unambiguously ascribed gender, two institutional types, and three major academic positions as major parameters.

## Methodological limitations and practical implications thus far

The use of academic age as a proxy for biological age raises a number of common questions. These pertain to the data set bias (usually Scopus or Web of Science, which come with their own linguistic, geographical, and disciplinary biases): other data sets than global bibliometric sources cannot be easily used in global interdisciplinary studies that go beyond single institutions or single countries. Furthermore, gender differences cannot be easily examined without massive gender ascription to publication authors. Not only do bibliometric data sets favor certain countries (Anglo-American) over other countries, but it is also that certain techniques (author disambiguation, author gender ascription) are much less reliable for certain parts of the world (South East Asia, including China) because of lower probability levels in ascribing individual names to gender and individual author IDs to real scientists and scholars. For example, gender ascription levels reach high percentages for countries such as Poland and Russia in contrast to China and South Korea (see Gender Probability Scores in Elsevier, 2020).

In addition, academic age can be used only for publishing scientists and scholars (all non-publishers are automatically excluded from analyses) and, in addition, scientists and scholars who publish in internationally indexed journals and predominantly in English. Globally, and particularly among countries that are new to the global science system, younger cohorts are generally more present in global data sets than older cohorts, with implications for age structure biases. Younger researchers tend to be more international as they have entered the research system in which different norms and practices matter—expectations to collaborate internationally are higher (Kwiek, 2018b). Both younger generations of researchers are more international—and the global science system, due to its complexity, is based on increased international collaboration (Rørstadt et al., 2021: 17–18; an overview of Europe, see Kwiek, 2021). However, if competing sources of global data are considered (e.g., the Google Scholar database), the bias in favor of relatively young scientists may be even higher (Radicchi & Castellano, 2013).

Furthermore, certain research questions can be examined much more adequately with large data sets so that academic ages are more averaged and much less adequately in individual cases or cases with a small number of observations (Nane et al. 2017: 726). The countries that have recently joined global science and the disciplines that continue to prioritize national languages and books and book chapters (rather than journal articles)—such as major segments of the humanities and arts—cannot be reliably examined with the use of academic age as a proxy for biological age; in this case, the age at which a doctoral degree was awarded may be a better methodological choice, data availability permitting (see Guns et al., 2019). However, access to dates of receiving the PhD degree on a massive national scale is technically possible only in several countries—those in which Current Research Information Systems (CRIS) or its variants are being implemented, usually in relation to national research assessment exercises (Italy, Norway, and Poland).

## Hypotheses of this research

Our research questions are linked to findings from previous studies that used both biological and academic ages as their variables and are embedded in the Polish national context. However, the research questions are also closely related to the data set at our disposal in which we have reliable data on every individual scientist and their gender, institutional affiliation, date of birth, date of obtaining doctorates, habilitations and professorial titles, as well as full metadata on their lifetime individual publication portfolios from Scopus (using Elsevier's ICSR Lab cloud computing services), including the date of the first publication.

We assume three hypotheses here. First, we assume that the biological age at which the first publication indexed in global databases like Scopus appears for a given scientist differs significantly for men and women: in practice, it may be lower for men (Hypothesis 1; confirmed). Second, we assume that the biological age at which a researcher publishes for the first time is significantly different in STEMM and non-STEMM disciplines (with a focus on the humanities and social sciences—the HUM and SOC fields): in practice, it may be lower for STEMM disciplines (Hypothesis 2; confirmed). Third, we assume that the biological age of a researcher at the time of first publication varies by institutional research intensity (particularly between a small group of research-intensive institutions on the one hand and all others on the other): in practice, it may be lower in institutions that are more heavily involved in research (Hypothesis 3; confirmed).

## Data and methods

### Data set

We used the integrated "Polish Science Observatory" database (see Kwiek & Roszka, 2021a: 4–6 for a more detailed description of its construction). Two large databases were merged: the data from an official national administrative and biographical register of all Polish scientists and scholars and the metadata from the Scopus publication and citation database for the decade of 2009–2018. The number of publications in the database was 158,743 articles and the number of unique authors was 25,463.

## Methods

### Gender classification scheme

There was no need to define the gender of Polish scientists and scholars through, for example, the various probabilistic methods, external data sets with gender-defined observations, or special gender defining software (for example, GenderizeR) (see Wais, 2016), as in our previous research that included the gender of international research collaborators of Polish scientists and scholars as one of the variables (Kwiek & Roszka, 2021a). In this case, gender is provided for all scientists and scholars with at least a PhD degree from all public science sectors present in our database, including 20,596 scientists and scholars in our final sample. It is a binary variable provided by the national registry of scientists. It was not possible to include any other options (based on "other" or "prefer not to disclose" responses) in the registry.

### Determining disciplines

It was necessary to determine the academic disciplines represented by Polish scientists and scholars. Every Polish scientist and scholar present in our integrated database was ascribed to one of 334 ASJC disciplines at the four-digit level and one of 27 ASJC disciplines at the two-digit level (following Abramo et al. (2020), who defined the Web of Science subject category for each Italian and Norwegian professor in their sample). In the ASJC system of disciplines used in Scopus, a journal publication can have one or multiple disciplinary classifications derived from journal disciplinary classifications. The dominant ASJC for each scientist was determined on the basis of all publications (type: article) present in their individual publication portfolios for the period 2009–2018 (the mode being the most frequently occurring value). In the case when there was no single mode, the dominant ASJC was randomly selected. Consequently, we had Polish scientists and scholars unambiguously defined by their gender and ASJC discipline, along with the metadata of all their publications indexed in the Scopus database. Three disciplines at the two-digit level were excluded from further analysis as they did not meet an arbitrary minimum threshold of 50 scientists per discipline (MULTI, NEURO, and NURS); consequently, our study is effectively based on 24 disciplines (16 belonging to STEMM and 8 belonging to non-STEMM fields). The list of ASJC disciplines is shown in Table 1 and they are described in the section Variables.

### Determining academic age

Having an individual scientist as the unit of analysis, we calculated the academic age based on the year of the first publication for every scientist in our sample. Our data set provides the date of birth and the dates of receiving all the relevant scientific degrees (doctoral degree for all and habilitation degree and professorship title for some scientists and scholars, depending on the stage of academic career).

The academic age we used ranges from 0 to 40 (as we generally assumed to include scientists and scholars in the biological age range of 30–70 years and with doctorates). We obtained the dates of the first publication indexed in the Scopus database for every Polish scientist using the application programming interface (API) protocol (a set of programming codes that enables data transmission between one software product and another) provided by Scopus. To use API, we applied individual Scopus IDs for every scientist in the sample. Our reference year was 2017 and the difference between 2017 and the year of the first publication was used to indicate the academic age as of 2017.

**Table 1** Descriptive statistics of biological and career age by gender, academic position, institutional type, and academic discipline (N=20,596)

| | Biological age | | | Career age | | |
|---|---|---|---|---|---|---|
| | Mean | SD | Median | Mean | SD | Median |
| *Gender* | | | | | | |
| Total | 47.0 | 10.6 | 45.0 | 43.2 | 9.3 | 41.0 |
| Female scientists | 45.1 | 9.7 | 43.0 | 41.2 | 8.2 | 39.0 |
| Male scientists | 48.4 | 11.0 | 46.0 | 44.6 | 9.9 | 42.0 |
| *Academic position* | | | | | | |
| Assistant Professor | 41.3 | 7.8 | 40.0 | 39.0 | 6.2 | 38.0 |
| Associate Professor | 50.8 | 8.4 | 49.0 | 45.6 | 8.4 | 45.0 |
| Full Professoror | 61.3 | 6.8 | 63.0 | 54.4 | 10.3 | 56.0 |
| *IDUB* | | | | | | |
| IDUB | 46.6 | 10.8 | 44.0 | 43.9 | 9.7 | 42.0 |
| Rest | 47.2 | 10.5 | 45.0 | 42.9 | 9.2 | 41.0 |
| *ASJC discipline* | | | | | | |
| AGRI | 47.6 | 10.4 | 46.0 | 42.0 | 7.9 | 40.0 |
| BIO | 45.3 | 10.3 | 43.0 | 44.6 | 9.3 | 43.0 |
| BUS | 46.3 | 9.7 | 44.0 | 36.3 | 6.6 | 34.0 |
| CHEM | 45.8 | 11.1 | 44.0 | 46.1 | 9.8 | 45.0 |
| CHEMENG | 47.3 | 11.7 | 45.0 | 43.7 | 10.4 | 41.0 |
| COMP | 46.5 | 10.5 | 44.0 | 44.1 | 9.0 | 42.0 |
| DEC | 48.1 | 11.2 | 43.5 | 41.6 | 11.8 | 37.5 |
| DENT | 45.0 | 9.8 | 45.0 | 39.9 | 8.8 | 39.0 |
| EARTH | 48.2 | 11.2 | 46.0 | 43.8 | 10.1 | 42.0 |
| ECON | 44.3 | 9.2 | 42.0 | 36.0 | 5.7 | 35.0 |
| ENER | 46.9 | 12.0 | 44.0 | 39.8 | 9.0 | 37.0 |
| ENG | 47.7 | 11.3 | 45.0 | 43.4 | 9.4 | 41.0 |
| ENVIR | 47.1 | 10.2 | 45.0 | 42.1 | 8.1 | 40.0 |
| HEALTH | 48.8 | 10.4 | 49.0 | 38.1 | 4.8 | 37.0 |
| HUM | 46.7 | 9.6 | 45.0 | 36.2 | 6.3 | 34.0 |
| IMMU | 45.7 | 9.3 | 45.0 | 45.7 | 8.1 | 45.0 |
| MATER | 46.6 | 10.9 | 44.0 | 45.4 | 9.6 | 43.0 |
| MATH | 47.1 | 11.3 | 45.0 | 45.5 | 9.8 | 43.0 |
| MED | 47.9 | 9.8 | 47.0 | 45.1 | 8.9 | 44.0 |
| PHARM | 44.5 | 10.3 | 42.0 | 43.9 | 9.1 | 42.0 |
| PHYS | 48.2 | 11.5 | 46.0 | 48.6 | 10.4 | 46.0 |
| PSYCH | 44.6 | 10.5 | 41.5 | 37.0 | 6.2 | 36.0 |
| SOC | 45.4 | 9.6 | 44.0 | 36.7 | 6.5 | 35.0 |
| VET | 47.3 | 10.4 | 46.0 | 44.5 | 7.8 | 43.5 |

For example, regardless of the biological age as stated in our data set and derived from a national registry of scientists, if a scientist had their first Scopus-indexed publication in 1997, their academic age in 2017 was 20 years; moreover, if they had their first publication in 2016, their academic age in 2017 was 1 year, thereby implying being at the very beginning of the academic career as defined through publications. It can be roughly assumed that in the natural sciences, the first publication in the database usually occurs during the last years of writing a doctorate—that is, around the age of 30 years in the Polish case (GUS, 2020). Thus, in a large simplification, the academic age of 0 would imply about 30 years of biological age, and the academic age of 20 would imply about 50 years of biological age. We also link academic age to the idea of career age as the age spent in science (only twice: in Fig. 1 and Table 1). In this sense, in the Polish context, a lifetime academic career, rare as it is, would include 40 years of life: an academic age in the range of 0–40 years (implying a career age of 30–70 years) and a biological age in the range of 30–70 years.

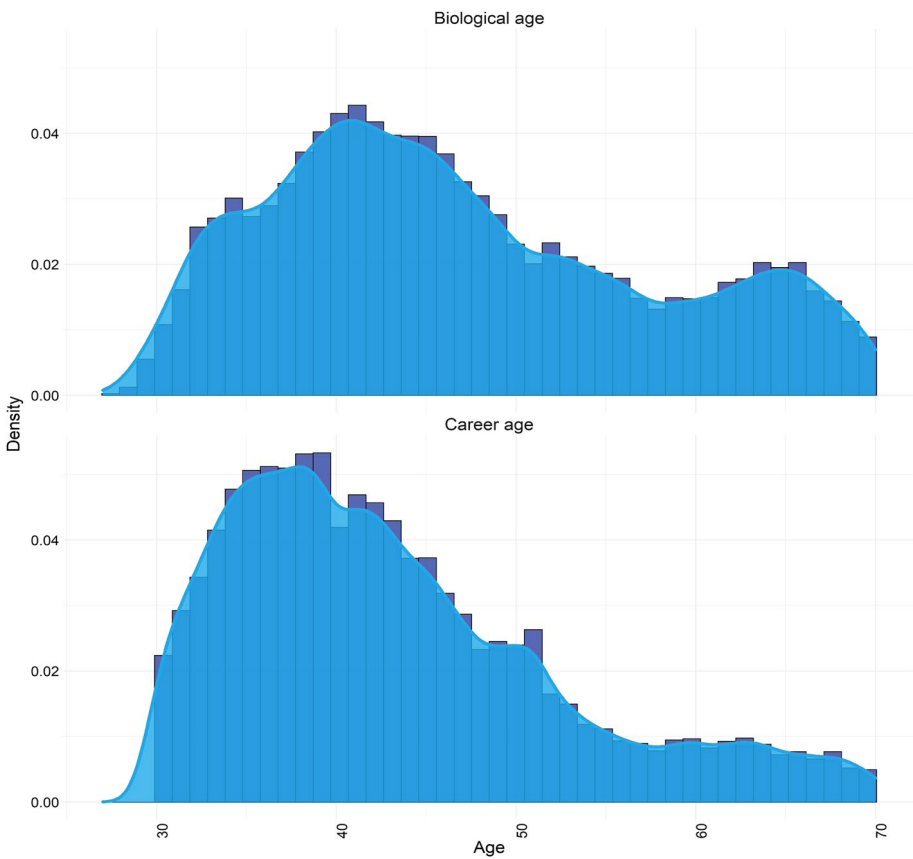

**Fig. 1** Distribution of biological age (top) and career age (bottom) and the Kernel density plot for all academic fields combined ($N = 20,596$)

## Sample

We began with a set of 25,463 scientists and scholars (14,886 male scientists and 10,577 female scientists—58.5% and 41.5%, respectively). This set included all scientists and scholars who were full-time employed in the higher education sector and who had at least a single article indexed in the Scopus database in the period 2009–2018 and who had at least a doctoral degree. Generally, 95%–97% of scientists and scholars in higher education have been employed full-time in the previous decade (GUS 2021, Table 1/44 in Electronic Annex). The sample includes all internationally visible (through publication type: article and in an international database type: Scopus) Polish academic scientists and scholars.

We defined the sample for this research in the following steps: we began with all scientists and scholars from all public science sectors present in our "Observatory" data set—99,935; then, we moved on to scientists and scholars with doctorates, which limited the number of our observations to 70,272; then, we moved on to scientists and scholars with doctorates and employed in higher education—54,448; among these, there were 32,937 scientists and scholars with publications of any type in Scopus in the period 2009–2018. Finally, there were 25,463 scientists and scholars with journal articles only.

We decided to include information on the entire academic production of individual authors in the database thus created (the original database contained bibliometric information only for the decade 2009–2018, and the reference period for the demographic and professional data of the authors was November 2017); in particular, we decided to include the year of the first publication. Such information was obtained from the Scopus API database for 21,285 individuals. Moreover, excluding individuals older than 70 years of age from the database and those for whom an incorrect date of first publication was found led to a reduction of the effective sample size to 20,596 authors ($N = 20,596$).

Figure 1 depicts the distribution of the sample by biological age (upper panel) and career age (bottom panel) for illustrative purposes: the Polish academic profession is relatively young, with the majority of scientists and scholars in the 30–45 age group and a substantial proportion of scientists and scholars who are older than 50 years of age (and older than 60 years of age). The distribution of the sample in terms of career age (derived from academic age to make the comparison clearer) looks similar: the majority of scientists and scholars have been publishing for no more than 15 years. The long tail on the right shows that although the proportion of older scientists and scholars is substantial (by biological age), they have been publishing internationally for a much shorter duration than their biological age indicates. Only a small minority of scientists and scholars (all academic disciplines combined) have been publishing for 25, 30, or 40 years. The generational difference between younger and older scientists and scholars and their publishing is clearly shown by the displacement of the sample to the left along the x-axis.

The descriptive statistics of the biological and career ages (used here for the sake of clarity and calculated as the academic age plus 30, adequately in the Polish case) of the sample are provided in Table 1, with major parameters such as gender, academic position (assistant, associate, and full professors), institutional type (research-intensive IDUB type and the rest), and ASJC academic disciplines. Career age here and below is used for the purpose of clarity only.

Several aspects are required to be emphasized here: the biological median age of 45 years is higher than the median career age of 41 years, and both are higher for male scientists than for female scientists. The difference between the two ages increases with subsequent academic positions: while for assistant professors it is small, for full professors it is substantial (two and seven years, respectively). This implies that the cohort of junior scientists began publishing internationally earlier than the cohort of senior scientists and particularly earlier than current full professors. In addition, the cross-disciplinary differentiation is clear: for STEMM disciplines, the discrepancies between median biological age and median career age are much lower than that for non-STEMM disciplines. Scientists in STEMM disciplines begin publishing internationally much earlier than scholars in non-STEMM disciplines. In major STEMM disciplines such as BIO (biochemistry, genetics, and molecular biology) there is no discrepancy at all and in CHEM, COMP, MATH, PHYS, and MED, the discrepancy is in the range of merely 1–2 years. In contrast, in such major non-STEM disciplines as BUS (business, management, and accounting), ECON (economics, econometrics, and finance), HUM (arts and humanities), and SOC (social sciences), the discrepancy reaches 8–10 years.

Clearly, for major social science and humanities fields, the discrepancy is fundamentally higher than for major natural science fields, which clearly indicates that for the currently employed academic workforce, biological age can be inferred for STEMM disciplines with marginal error; however, for non-STEMM disciplines, the error is substantial and allocating scholars to major career stages—beginning, early, middle, and late—can be misleading.

The gender distribution of our sample of Scopus publishers is approximately 40/60 (42.5% of female scientists and 57.5% of male scientists) and is similar to the current distribution of higher education personnel (47.6% and 52.4% in 2020) (GUS 2021: 126). Approximately half the scientists and scholars in our sample are middle-aged (in the 40–54 age bracket; 47.3%) and over half of them are assistant professors (55.7.0%, see Table 2).

## Variables

We used numerical and categorical variables which were of a biological, demographic, administrative, and institutional nature. Biological age in our database was provided by the national registry of scientists ($N=99,935$) and age in full years as of 2017 was used. We used three major age groups: young (39 and younger; $N=5708$), middle-aged (40–54; $N=9746$), and older (55 and older; $N=5142$) scientists. Gender was provided by the national registry of scientists "The Polish Science" (*Nauka Polska, N=99,935*).

Three Polish academic degrees were used as proxies of internationally comparable academic positions: doctoral degree only (as a proxy of assistant professor; $N=11,470$); habilitation or postdoctoral degree (as a proxy of associate professor; $N=6242$); and professorship title (as a proxy of full professor; $N=2884$). All scientists and scholars without doctoral degrees and with institutional affiliations outside of the higher education sector were excluded from the analysis. Four career stages based on academic experience were used: beginning ($B$, <5 years of academic experience—since the first publication in Scopus), early career ($E$, 5–14 years), middle career ($M$, 15–29 years), and late career ($L$, 30 years and more).

All scientists and scholars were ascribed to one of 24 Scopus ASJC disciplines (as three disciplines were excluded from the analysis because the number of scientists ascribed to them was lower than 50) and individually-determined dominant disciplines were used. STEMM disciplines are the following: AGRI (agricultural and biological sciences); BIO (biochemistry, genetics, and molecular biology); CHEMENG (chemical engineering); CHEM (chemistry); COMP (computer science); DEC (decision science); EARTH (Earth and planetary sciences); ENER (energy); ENG (engineering); ENVIR (environmental science); IMMU (immunology and microbiology); MATER (materials science); MATH

**Table 2** Structure of the sample, all Polish internationally visible university professors, by gender, age group, academic position, and discipline (Young scientists indicate those aged 39 and younger, Middle-aged indicates those aged 40–54, and Older indicates those aged 55 and over)

| | Female | | | Male | | | Total | | |
|---|---|---|---|---|---|---|---|---|---|
| | N | % col | % row | N | % col | % row | N | % col | % row |
| *Age group* | | | | | | | | | |
| Young (39 and younger) | 2828 | 32.3 | 49.5 | 2880 | 24.3 | 50.5 | 5708 | 27.7 | 100.0 |
| Middle-aged (40–54) | 4368 | 49.9 | 44.8 | 5378 | 45.4 | 55.2 | 9746 | 47.3 | 100.0 |
| Older (55 and older) | 1564 | 17.9 | 30.4 | 3578 | 30.2 | 69.6 | 5142 | 25.0 | 100.0 |
| **Total** | **8760** | **100.0** | **42.5** | **11,836** | **100.0** | **57.5** | **20,596** | **100.0** | **100.0** |
| *Academic position* | | | | | | | | | |
| Assistant Professor | 5557 | 63.4 | 48.4 | 5913 | 50.0 | 51.6 | 11,470 | 55.7 | 100.0 |
| Associate Professor | 2443 | 27.9 | 39.1 | 3799 | 32.1 | 60.9 | 6242 | 30.3 | 100.0 |
| Full Professor | 760 | 8.7 | 26.4 | 2124 | 17.9 | 73.6 | 2884 | 14.0 | 100.0 |
| **Total** | **8760** | **100.0** | **42.5** | **11,836** | **100.0** | **57.5** | **20,596** | **100.0** | **100.0** |
| *Discipline (ASJC) – STEMM* | | | | | | | | | |
| AGRI | 1215 | 13.9 | 54.0 | 1037 | 8.8 | 46.0 | 2252 | 10.9 | 100.0 |
| BIO | 900 | 10.3 | 61.7 | 558 | 4.7 | 38.3 | 1458 | 7.1 | 100.0 |
| CHEM | 612 | 7.0 | 52.1 | 562 | 4.7 | 47.9 | 1174 | 5.7 | 100.0 |
| CHEMENG | 155 | 1.8 | 40.6 | 227 | 1.9 | 59.4 | 382 | 1.9 | 100.0 |
| COMP | 141 | 1.6 | 17.1 | 684 | 5.8 | 82.9 | 825 | 4.0 | 100.0 |
| DEC | 20 | 0.2 | 43.5 | 26 | 0.2 | 56.5 | 46 | 0.2 | 100.0 |
| EARTH | 333 | 3.8 | 34.9 | 622 | 5.3 | 65.1 | 955 | 4.6 | 100.0 |
| ENER | 60 | 0.7 | 26.0 | 171 | 1.4 | 74.0 | 231 | 1.1 | 100.0 |
| ENG | 412 | 4.7 | 15.0 | 2336 | 19.7 | 85.0 | 2748 | 13.3 | 100.0 |
| ENVIR | 735 | 8.4 | 51.9 | 680 | 5.7 | 48.1 | 1415 | 6.9 | 100.0 |
| IMMU | 72 | 0.8 | 75.8 | 23 | 0.2 | 24.2 | 95 | 0.5 | 100.0 |
| MATER | 417 | 4.8 | 34.2 | 801 | 6.8 | 65.8 | 1218 | 5.9 | 100.0 |
| MATH | 213 | 2.4 | 26.4 | 595 | 5.0 | 73.6 | 808 | 3.9 | 100.0 |
| MED | 1620 | 18.5 | 55.2 | 1314 | 11.1 | 44.8 | 2934 | 14.2 | 100.0 |
| PHARM | 148 | 1.7 | 69.8 | 64 | 0.5 | 30.2 | 212 | 1.0 | 100.0 |
| PHYS | 147 | 1.7 | 17.6 | 690 | 5.8 | 82.4 | 837 | 4.1 | 100.0 |
| *Discipline (ASJC) – non-STEMM* | | | | | | | | | |
| BUS | 280 | 3.2 | 53.0 | 248 | 2.1 | 47.0 | 528 | 2.6 | 100.0 |
| DENT | 48 | 0.5 | 73.8 | 17 | 0.1 | 26.2 | 65 | 0.3 | 100.0 |
| ECON | 143 | 1.6 | 50.4 | 141 | 1.2 | 49.6 | 284 | 1.4 | 100.0 |
| HEALTH | 20 | 0.2 | 37.7 | 33 | 0.3 | 62.3 | 53 | 0.3 | 100.0 |
| HUM | 405 | 4.6 | 51.0 | 389 | 3.3 | 49.0 | 794 | 3.9 | 100.0 |
| PSYCH | 148 | 1.7 | 62.7 | 88 | 0.7 | 37.3 | 236 | 1.1 | 100.0 |
| SOC | 381 | 4.3 | 50.8 | 369 | 3.1 | 49.2 | 750 | 3.6 | 100.0 |
| VET | 135 | 1.5 | 45.6 | 161 | 1.4 | 54.4 | 296 | 1.4 | 100.0 |
| **Total** | **8760** | **100.0** | **42.5** | **11,836** | **100.0** | **57.5** | **20,596** | **100.0** | **100.0** |

Total by gender for each category (row and column percentages) in bold

(mathematics); PHARM (pharmacology, toxicology, and pharmaceutics); PHYS (physics and astronomy); and MED (medicine). Non-STEMM disciplines are the following: BUS (business, management, and accounting); DENT (dentistry), ECON (economics, econometrics, and finance); HEALTH (health professions); HUM (arts and humanities); PSYCH (psychology); SOC (social sciences); and VET (veterinary). Research-intensive institutions are the ten institutions (from among 85 examined) selected in 2019 for the IDUB (or "Excellence Initiative–Research University") national program.

## Results

### Example

Before going further, we provide an example: the first publication of researcher A (a male in the ECON discipline—economics, econometrics, and finance who obtained his PhD in 1995, his habilitation in 2005, and his professorship in 2012) was, according to the Scopus database, published in 2015. The following assumption (based on selected global papers discussed above) is made in this regard: if 2015 is the beginning of an academic career for this scientist in the sense of beginning to publish (or becoming an active member of the academic profession), then his academic age is 0. Therefore, in 2017, he should have a biological age of 32. However, our "Observatory" database using his date of birth indicates 52 years as his biological age in 2017.

Thus, an inconsistency arises: using academic age as a proxy for biological age, scientist A is young and just beginning his scientific career (academic age = 2, career stage: beginning); whereas using biological age, we conclusively find that scientist A is older (52 years old, career stage: middle). Our task below is to estimate this mismatch between academic age and biological age on a large scale of the entire national science system, depending on selected parameters.

### Correlation of biological and academic ages

In order to analyze the relationship between biological age (in the range of 30–70 years) on academic age (in the range of 0–40 years), a linear correlation analysis between these variables was performed (Table 3). Regardless of the approach of the correlations presented, a positive relationship was observed in each case. All Pearson's linear correlation coefficients are significantly different from zero (at a significance level of $\alpha = 0.05$). Another clearly visible pattern is the usually strong or very strong correlation observed for disciplines belonging to the STEMM cluster. For most STEMM disciplines, a stronger correlation was observed than for all disciplines together (i.e., Total). The scatter plots (Fig. 2) clearly indicate that the vast majority of individuals publishing in STEMM disciplines publish their first article at a relatively young age (the points shift to the right on the X axis). Moreover, the highest correlations were observed for STEMM disciplines such as chemistry (CHEM, $r = 0.889$), physics and astronomy (PHYS, $r = 0.883$), mathematics (MATH, $r = 0.849$) or medicine (MED, $r = 0.750$).

In contrast, disciplines such as business (BUS), management and accounting, arts and humanities (HUM), psychology (PSYCH), and social science (SOC) (representing the non-STEMM field)—relatively abundantly represented by Polish scholars—are characterized by a relatively low strength of the relationship between biological and academic ages. The value of the correlation coefficient does not exceed 0.5 in any case and remains in the range from 0.354 for humanities to 0.488 for psychology; the scatter analysis (Fig. 2) indicates a clear shift of points to the left on the X axis.

**Table 3** Pearson's correlation coefficients for Polish scientists and scholars between biological age and academic age and test for association between paired samples statistics in terms of sex, institutional type (research-intensive IDUB institutions and the rest), academic position and ASJC discipline ($N = 20,596$)

| Category | Estimate | t-statistic | p-value | Df | Confidence Interval – LB | Confidence Interval – UB |
|---|---|---|---|---|---|---|
| Total | 0.691 | 137.1 | <0.001 | 20,594 | 0.684 | 0.698 |
| Female scientists | 0.655 | 81.2 | <0.001 | 8758 | 0.643 | 0.667 |
| Male scientists | 0.697 | 105.8 | <0.001 | 11,834 | 0.688 | 0.706 |
| Institution – Rest | 0.670 | 106.6 | <0.001 | 13,961 | 0.661 | 0.679 |
| Institution – IDUB | 0.739 | 89.4 | <0.001 | 6631 | 0.728 | 0.750 |
| Associate Professor | 0.509 | 46.7 | <0.001 | 6240 | 0.490 | 0.527 |
| Full Professor | 0.339 | 19.4 | <0.001 | 2882 | 0.307 | 0.371 |
| Assistant Professor | 0.583 | 76.9 | <0.001 | 11,468 | 0.571 | 0.595 |
| IMMU | 0.903 | 20.3 | <0.001 | 93 | 0.858 | 0.935 |
| CHEM | 0.889 | 66.5 | <0.001 | 1172 | 0.876 | 0.900 |
| PHYS | 0.883 | 54.5 | <0.001 | 835 | 0.868 | 0.898 |
| PHARM | 0.856 | 24.0 | <0.001 | 210 | 0.816 | 0.889 |
| MATH | 0.849 | 45.6 | <0.001 | 806 | 0.828 | 0.867 |
| MATER | 0.836 | 53.0 | <0.001 | 1216 | 0.818 | 0.852 |
| BIO | 0.820 | 54.7 | <0.001 | 1456 | 0.803 | 0.836 |
| DENT | 0.789 | 10.2 | <0.001 | 63 | 0.675 | 0.866 |
| VET | 0.773 | 20.9 | <0.001 | 294 | 0.723 | 0.815 |
| MED | 0.750 | 61.4 | <0.001 | 2932 | 0.734 | 0.766 |
| DEC | 0.749 | 7.5 | <0.001 | 44 | 0.587 | 0.854 |
| EARTH | 0.746 | 34.5 | <0.001 | 953 | 0.716 | 0.772 |
| CHEMENG | 0.729 | 20.8 | <0.001 | 380 | 0.679 | 0.773 |
| COMP | 0.721 | 29.9 | <0.001 | 823 | 0.687 | 0.753 |
| ENG | 0.707 | 52.4 | <0.001 | 2746 | 0.688 | 0.725 |
| ENVIR | 0.650 | 32.1 | <0.001 | 1413 | 0.618 | 0.679 |
| AGRI | 0.612 | 36.7 | <0.001 | 2250 | 0.586 | 0.637 |
| ENER | 0.568 | 10.4 | <0.001 | 229 | 0.473 | 0.649 |
| HEALTH | 0.547 | 4.7 | <0.001 | 51 | 0.324 | 0.712 |
| PSYCH | 0.488 | 8.5 | <0.001 | 234 | 0.384 | 0.579 |
| SOC | 0.464 | 14.3 | <0.001 | 748 | 0.406 | 0.519 |
| BUS | 0.441 | 11.3 | <0.001 | 526 | 0.369 | 0.507 |
| ECON | 0.428 | 7.9 | <0.001 | 282 | 0.328 | 0.518 |
| HUM | 0.354 | 10.7 | <0.001 | 792 | 0.292 | 0.414 |

Thus, the results reveal that most representatives of non-STEMM disciplines, despite their relatively advanced biological age, are relatively young in terms of their academic age. Their first publication in the Scopus database—that is, their publication debut in the global scientific arena—clearly begins later than in the case of STEMM scientists.

This observation is consistent with common intuitions and previous surveys (Kwiek, 2015a, 2020), according to which it takes a shorter amount of time for representatives of the sciences in Poland to participate in the global circulation of science than for representatives of the humanities, social sciences, or economics (among other reasons due to the expansion of the private sector between 1990 and 2006 on which much of the energy of non-STEMM scholars had been focused, Kwiek & Szadkowski, 2019).

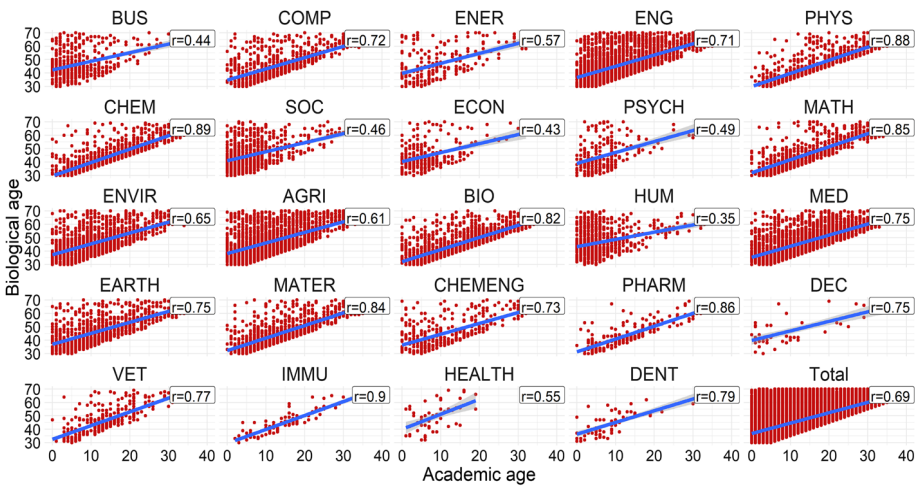

**Fig. 2** Scatter analysis for Polish scientists and scholars between academic age (range 0–40 years) and biological age (range 30–70 years) by ASJC discipline ($N = 20,596$)

Smaller differences in correlations were observed for the other independent variables: university type in terms of research intensity (research-intensive IDUB institutions vs. the rest), academic position, and gender. For faculty at research-intensive universities (IDUB), the correlation is higher than for IDUB ($r = 0.739$ vs. $r = 0.670$), but for both, the correlation coefficient is close to the correlation for all observations ($r = 0.691$). Moreover, the correlation is higher for men than that for women ($r = 0.697$ vs. $r = 0.655$).

We have not focused in this paper on the year scientists and scholars have been granted their doctoral degrees as a proxy for an academic age for a simple reason: although in our dataset we have the date for every person, this biographical attribute is rarely available on a national scale in other systems. Our idea was to assess a proxy that is widely available through large bibliometric databases like Scopus and therefore our analyses are performed on the date of first publication. However, the results of a linear correlation analysis between the age of earning a PhD and academic age show important differences compared with the above analyses (see Table 5 in the Data Appendix). While correlations for STEMM disciplines are high and correlations for non-STEMM are low, there are large differences between correlations for male and female scientists and correlations for research intensive IDUB institutions and the rest, not observable in above analyses. Somehow surprisingly, the correlation coefficient for female scientists is almost twice as high as the one for male scientists, and the correlation coefficient for research intensive IDUB institutions is twice as high as the one for the rest ($r = 0.676$ vs. $r = 0.375$; and $r = 0.752$ vs. $r = 0.371$, respectively). These differences may result from possibly higher publication requirements for doctoral degrees in research-intensive institutions and from possibly higher productivity of women in the early years of employment.

Furthermore, the differences in academic positions reveal (Fig. 3) that although the correlation coefficients are significantly lower, the differences in the slope of the regression curve clearly show that scientists and scholars in lower positions are generally academically younger than their colleagues in higher positions (shift of points to the left on the X-axis), which is probably due to the correlation of position with biological age. Simultaneously, as the academic seniority increases, the strength of the correlation clearly decreases (from *r* = 0.583 for assistant professors to only *r* = 0.339 for full professors).

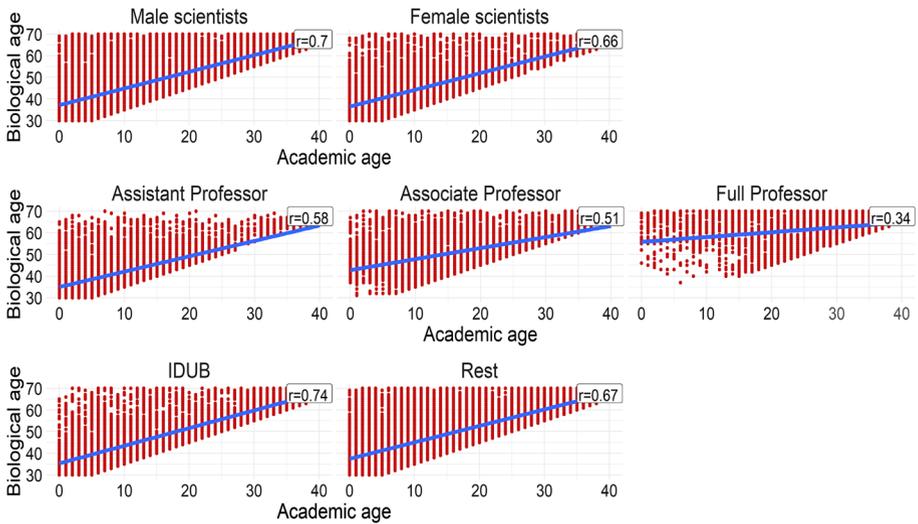

**Fig. 3** Scatter analysis for Polish scientists and scholars between academic age (range 0–40 years) and biological age (range 30–70 years) by gender (top panel) institutional type (research-intensive IDUB institutions vs. the rest, middle panel),and academic position (bottom panels) (*N* = 20,596)

These results clearly show the diversity of scientists and scholars in terms of when they begin to publish: current full professors, the oldest and highest in the academic hierarchy, are the latest to begin; and the youngest and at the beginning of their academic career, scientists only with a doctorate (assistant professors), find international publishing more natural. These findings confirm the results of earlier surveys and interviews that reveal radical generational differences in Polish science (Kwiek, 2015b) and survey results linking "internationalists" and "locals" in research to age and academic generations (Kwiek, 2020). Remarkably, an earlier survey research (conducted on 4000 returned questionnaires) is strongly corroborated by the detailed large-sample research presented here.

Intergenerational differences in international publishing are due to distinctly different starting conditions for different groups of researchers: different opportunities for international collaboration and different institutional requirements at successive stages of academic career development, particularly growing after 2010, when two series of structural higher education reforms (2010–2012 and 2016–2018) were initiated. Faculty at the level of assistant professors and, thus, also predominantly young faculty, have been embarking on their academic careers for a decade now, under radically better financial conditions and changed political and social realities than their colleagues who are currently associate or full professors. The inter-cohort differences clearly reveal the evolution of the Polish system of science.

Analyzing the distribution of biological age by individual academic ages for all scientists and scholars—regardless of discipline, institutional type, academic position, and

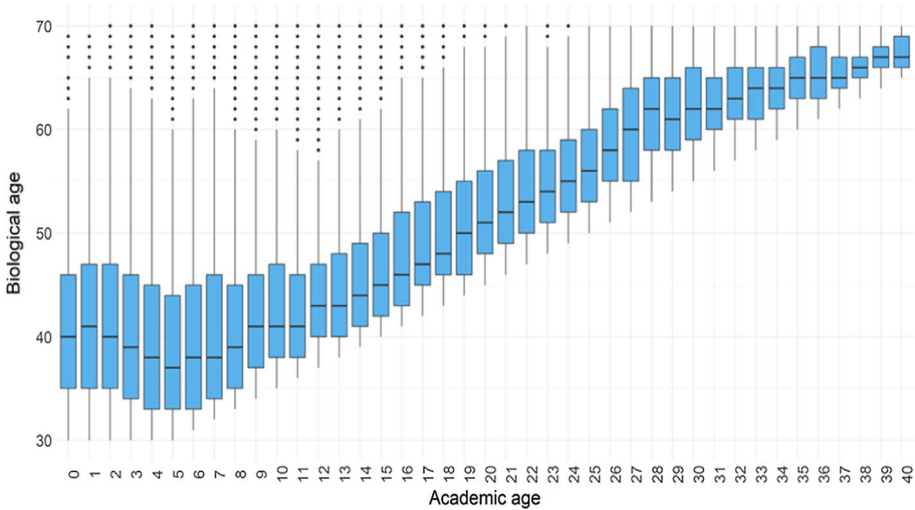

**Fig. 4** Distribution of biological age for Polish scientists and scholars for individual academic years of age (*N*=20,596)

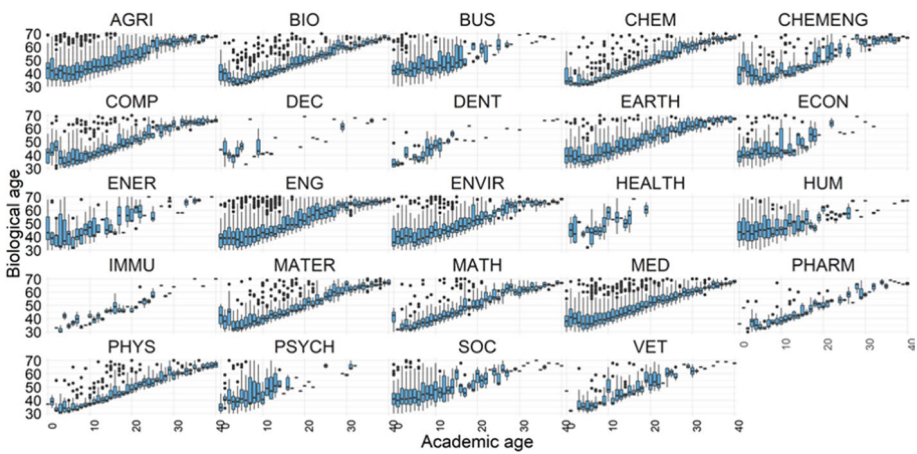

**Fig. 5** Distribution of biological age for Polish scientists and scholars for individual academic years of age by academic disciplines (*N*=20,596)

gender—there is an increasing but fading trend of the medians of biological age and a decreasing variation of biological age with increasing academic age (Fig. 4). For gender and institutional type (IDUB vs. rest), no significant differences noted, while disciplines (Fig. 5) and academic position (Fig. 6) have a strong influence on the shape of the biological age distribution.

For the STEMM disciplines, the trend of medians in many cases (e.g., BIO, CHEM, COMP, EARTH, ENG, ENVIR, MATER, MATH, MED, or PHYS) appears not to be fading but largely linear, once again highlighting the strong association between academic age and biological age in the STEMM field. Although academic careers evolve with the age of the scientist, for the non-STEMM field, it appears that the trend does not occur at all in certain cases (particularly in large disciplines such as ECON, HUM, or PSYCH) or is barely noticeable (as in BUS and SOC). Similarly, the variability in the distribution of biological age, which decreases with academic age for STEMM disciplines, appears to be unchanged or changing in a disordered manner. In terms of academic positions, the distribution of biological age for individual years of academic age appears to be similar to the general pattern (i.e., somewhat resembling a logistic curve—fading growth) for assistant professors and associate professors, but behaves rather differently for full professors. Among full professors, there is a clear dominance of scientists and scholars who are older in biological age for almost every year of academic age.

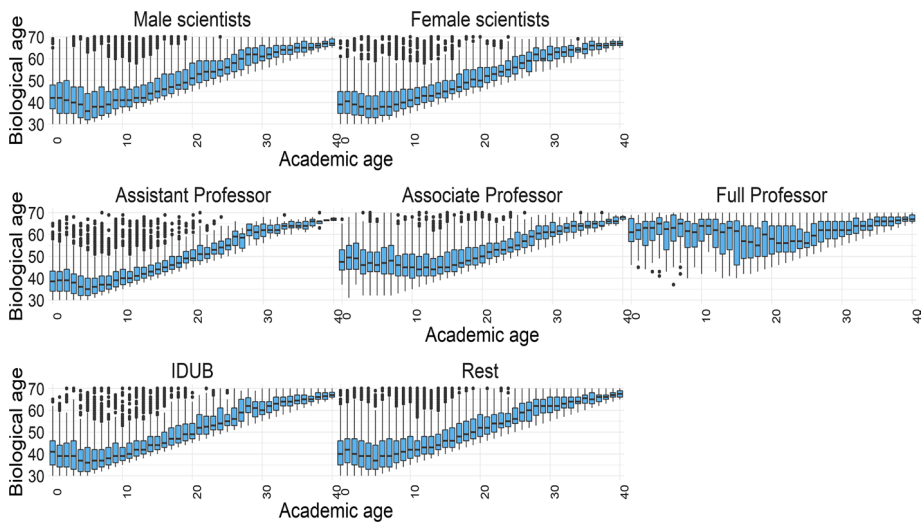

**Fig. 6** Distribution of biological age for Polish scientists for individual academic years of age by gender, academic position, and institutional research intensity (IDUB) ($N = 20,596$)

The interquartile ranges (Fig. 7) decrease as academic age increases, and they vary between 11 and 13 years for the low academic ages to 2–3 years for the oldest ages. In other words, variability decreases as academic age increases. While for the younger cohorts the deviation from the median age for the middle 50% of academics is approximately $\pm 6$ years, it is only $\pm 2$ years for the older cohorts.

## Contingency analysis

The conclusions obtained from the correlation analysis are confirmed by the contingency analysis of academic career stages (four stages) with academic position (three positions) and age groups (four age groups). We allocated all scientists and scholars to four career stages or years of academic publishing (academic age brackets): beginning (*B*, less than five years of academic experience since the first publication in Scopus), early career (*E*, 5–14 years), middle career (*M*, 15–29 years), and late career (*L*, 30 and over). In very approximate terms, if academic age is 0 years (beginning of publication career in the sense of first publication = 30 years), then those in the beginning stage are aged 30–34 years, those in the early career stage are aged 35–44 years, those in the middle career stage are aged 45–59 years, and those in the late career stage are aged

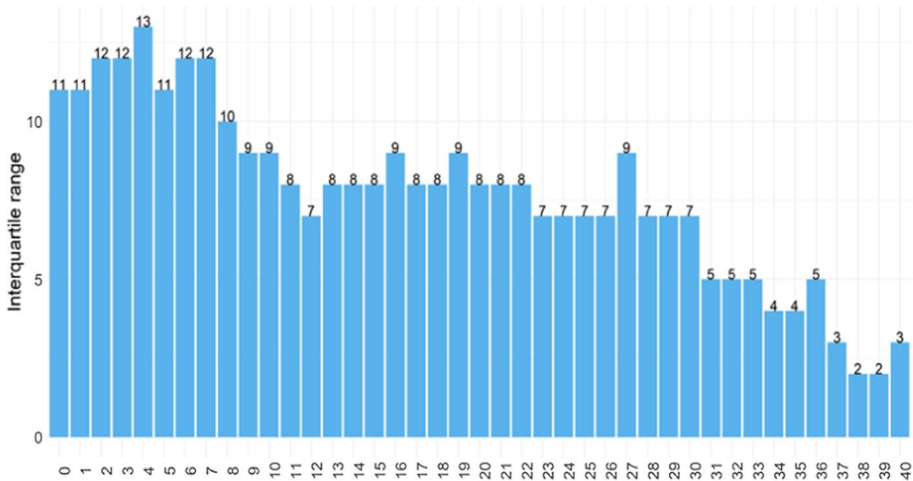

**Fig. 7** Biological age interquartile range for Polish scientists and scholars for individual academic years of age (range 0–40 years) ($N = 20,596$)

60 years and over. However, in all analyses, we use a well-defined and strictly determined academic age range of 0–40 years for each researcher.

An example of the differences between STEMM and non-STEMM disciplines is very well illustrated in the comparison between CHEM (chemistry) and HUM (arts and humanities) disciplines, both relatively populous (the upper panel in Fig. 8 presents age groups and the lower panel presents academic positions). In chemistry, the vast majority of assistant professors (62.5%) are in the early stage of their careers. On the other hand, a majority of associate professors (65.7%) are in the middle stage, while a majority of full professors (64.3%) are in the late stage. It is evident that the correlation of successive promotions and stages of scientific career with the advancement of own scientific work is understood as a bracket of academic age. Chemistry is an excellent example of a discipline in which academic positions correspond with the publication trajectories of scientists and scholars: full professors mostly publish much longer than other categories of scientists, as is expected. The contingency analyses by academic position (bottom panel) and by age group (top panel) are similar in this case, with the highest percentage of young scientists under the age of 40 (81.2%) being in the early career stage; this is consistent with the idea that younger generations entering the academic profession have been well internationalized in research.

In contrast, this division is not clear-cut in the humanities. We can observe a shift in the academic career stage (defined by the date of first publication) toward a younger age (Fig. 8, upper panel) and earlier academic career stage (Fig. 7, lower panel). Thus, among assistant professors at HUM, two-thirds (65.0%) are at the initial (beginning) stage of their career rather than the early stage, as was expected; but most importantly, the subsequent positions do not indicate a clear distinction among career stages—that is, one cannot identify a stage clearly dominated by associate professors or full professors. Scholars in these positions are mainly in the beginning and early stages of their careers, rather than in middle and late career stages, as suggested in the model (confirmed for CHEM above).

This implies that in the non-STEMM disciplines such as HUM arts and humanities (as in BUS business, management, and accounting; ECON economics, econometrics, and finance; PSYCH psychology; and SOC social sciences), Polish scholars began publishing their articles in the international circulation radically later than those in chemistry, which is compared here, but also in disciplines such as BIO and PHYS. It also implies that in the humanities, it is mainly the youngest scholars and those at the earliest stage of their scientific careers who publish internationally.

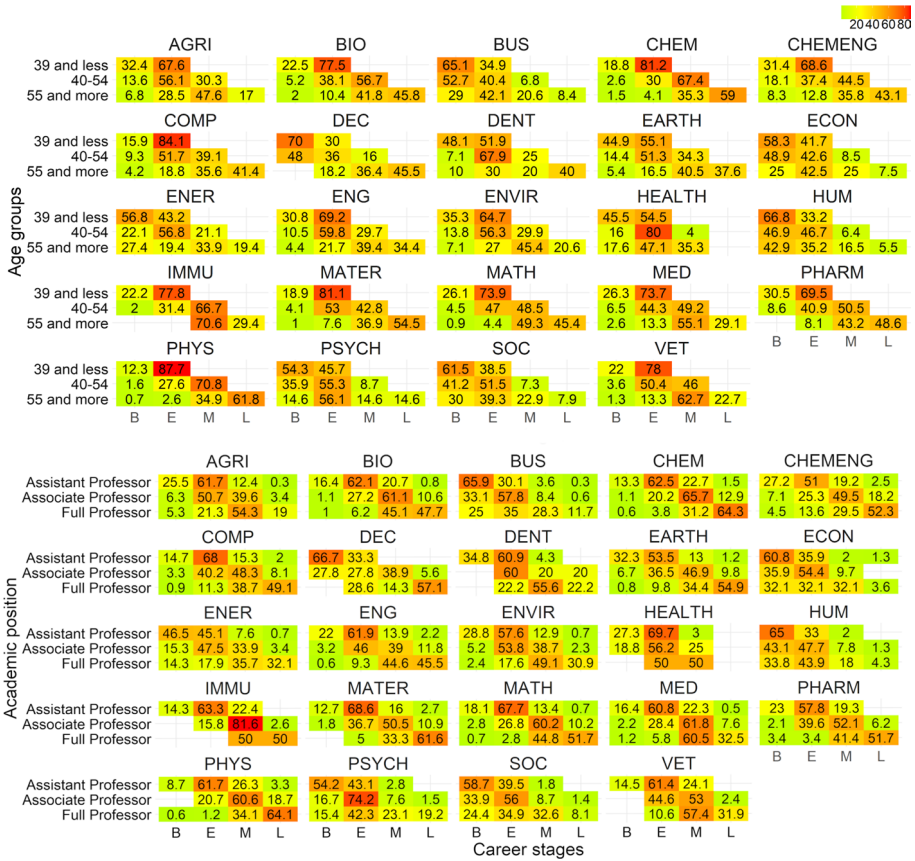

**Fig. 8** Contingency analysis of Polish scientists and scholars between biological age and academic age groups by academic position (upper panel) and ASJC discipline (lower panel). Four career stages are analyzed: *B*—beginning, *E*—early career, *M*—middle career, and *L*—late career. The contingency level increases as the colors change from green to red ($N=20{,}596$)

Analyzing the contingency between academic position and academic career stages (Fig. 8, lower panel), the evident pattern is that scholars in almost all non-STEMM disciplines are delayed by at least one academic career stage compared to STEMM disciplines. Moreover, in non-STEMM disciplines, associate professors and full professors do not dominate at all in the late stages of their careers, which implies that they publish internationally for shorter periods of time than one might assume.

An analogous pattern is observed for the contingency of age groups and academic careers (Fig. 8, upper panel): the STEMM disciplines clearly show advancement to later academic career groups with increasing age group, while for the non-STEMM disciplines this relationship is not obvious. We observe a significant shift over time for non-STEMM

disciplines—the first articles of non-STEMM scholars are published late in their careers, often only in the phase of working as a full professor. This has one implication: the academic age in the case of non-STEMM disciplines does not keep up with the biological age, and full professors are often at the same stage of academic age as assistant professors. Thus, the inference of biological age of scientists in STEMM disciplines proves to be an adequate approximation of reality, while the same inference in non-STEMM disciplines proves to lead to erroneous conclusions and distorted results.

The later date of the first publication for non-STEMM authors can be linked to both external and internal factors. External factors include, first, the differentiated representativeness of Polish research outputs in Scopus by discipline, with low representativeness of outputs in non-STEMM disciplines; and, second, the language coverage of Scopus, with a limited number of Polish-language journals. Weaker representation of non-STEMM journals compared with STEMM journals in Scopus (especially in social sciences, and even more so, in humanities) was often discussed in literature (see Aksnes & Sivertsen, 2019; Singh et al., 2021; Harzing, 2019). And internal factors include, first, historically consistently weaker focus on international publishing in social sciences and humanities, at least until the two waves of higher education reforms in 2010s, compared with strong focus on publishing in national journals (both prior to the collapse of the communist regime in 1989 and in the post-communist period); and, second, weaker focus on publishing in English, compared with strong focus on publishing in Polish. Poland has a shorter history in international academic publishing generally, with less research resources such as funding and infrastructures (see Mongeon & Hus, 2016). The reforms introducing new types of research assessment exercise in 2011 and 2017, with new internationally-oriented individual and institutional publishing requirements, slowly change the publishing practice of scientists and scholars.

Researchers in social sciences and humanities represent high levels of what was termed "multilingual publishing": a recent study shows that 8.1 percent of them publish only in English (the lowest rate in a sample of 7 European countries); and more importantly, 48.3 percent of them publish only in Polish (the highest rate). In more general terms, 44.1 percent of researchers publish in English and 88.4 percent publish in Polish (Kulczycki et al. 2020: 1375). However, publication patterns in social sciences and humanities in the Central and Eastern European countries are becoming increasingly similar to those in Western Europe and the Nordics, as a recent study of five European countries shows (Petr et al., 2021). As a national case study of Norway highlights, Scopus covers 89 percent of the total Norwegian scientific and scholarly output in medicine and health and 85 percent in natural sciences and technology, compared with merely 48 percent in social sciences and 27 percent in humanities (Aksnes & Sivertsen, 2019: 1). The later date of the first publication in non-STEMM disciplines makes inferring biological age from individual publication histories much less reliable—but the factors are both external (Scopus-related) and internal (Polish publishing patterns).

## Modeling approach: linear regression model

The correlation analysis of academic age and biological age with particular independent variables in the disjointed (two-dimensional) approach presented above leads to the observation of interesting patterns, but only the combined (multivariate) impact of all of the variables on biological age enables the provision of a full picture of the studied phenomenon.

In order to conduct multivariate analysis, a linear multivariate regression model was created, where the dependent variable was biological age and the independent variables were (1) academic age, (2) gender, (3) institutional type, (4) academic discipline, and (5) academic position. The resulting model explained 61.5% of the variation in biological age and the standard error was 6.59 years (i.e., when determining an age from the estimated model, we are off the mark by an average of 6.59 years; the relative error was 14%). All interpretations are subject to the *ceteris paribus* assumption and the significance level used was $\alpha = 0.05$ (Table 4).

According to the model, if academic age increases by one year, biological age increases by 0.6 years on average. This is also the variable with the strongest influence on the dependent variable (the standardized coefficient is 0.523, which is the largest among all independent variables). The characteristic with the second strongest effect is academic position. The position of full professor has a positive impact compared to the position of assistant professor (reference category; the increase in age is on average 11 years greater); simultaneously, the position of associate professor is also characterized by a high impact (increase on average 5.4 years). This relationship applies only to the studied population of scientists and scholars publishing in the Scopus database, and not all Polish scientists and scholars. Moreover, the predictor associated with working at universities other than the 10 research universities (reference category) positively influences biological age (on average by 1.3 years). Importantly, gender does not significantly affect the prediction of biological age.

**Table 4** Linear model coefficients (dependent variable: biological age; reference categories: gender—female, institutional type: research-intensive IDUB, ASJC discipline—HUM, and academic position—assistant professor) ($N = 20,596$)

| $R^2 = 0.615$, SE $= 6.59$ | Estimate | Standardized | SE | $t$-value | Pr($> |t|$) | VIF |
|---|---|---|---|---|---|---|
| (Intercept) | 38.269 | – | 0.250 | 152.998 | < 0.001 | – |
| Academic age | 0.595 | 0.523 | 0.007 | 90.076 | < 0.001 | 1.867 |
| Gender: Male scientists | 0.094 | 0.004 | 0.101 | 0.934 | 0.350 | 1.179 |
| Institutional research intensity: Rest | 1.253 | 0.055 | 0.103 | 12.206 | < 0.001 | 1.089 |
| ASJC discipline: AGRI | −2.460 | −0.072 | 0.278 | −8.846 | < 0.001 | 1.585 |
| ASJC discipline: BIO | −5.352 | −0.129 | 0.299 | −17.876 | < 0.001 | 1.226 |
| ASJC discipline: BUS | 0.290 | 0.004 | 0.371 | 0.780 | 0.436 | 1.512 |
| ASJC discipline: CHEM | −5.983 | −0.131 | 0.312 | −19.182 | < 0.001 | 1.159 |
| ASJC discipline: CHEMENG | −2.709 | −0.034 | 0.416 | −6.517 | 0.000 | 1.383 |
| ASJC discipline: COMP | −3.768 | −0.070 | 0.336 | −11.231 | < 0.001 | 1.018 |
| ASJC discipline: DEC | −1.910 | −0.008 | 1.001 | −1.909 | 0.056 | 1.027 |
| ASJC discipline: DENT | −2.616 | −0.014 | 0.853 | −3.068 | 0.002 | 1.417 |
| ASJC discipline: EARTH | −2.118 | −0.042 | 0.322 | −6.576 | 0.000 | 1.123 |
| ASJC discipline: ECON | −1.724 | −0.019 | 0.457 | −3.775 | 0.000 | 1.102 |
| ASJC discipline: ENER | −0.818 | −0.008 | 0.495 | −1.653 | 0.098 | 2.086 |
| ASJC discipline: ENG | −2.162 | −0.069 | 0.275 | −7.875 | 0.000 | 1.521 |
| ASJC discipline: ENVIR | −2.515 | −0.060 | 0.297 | −8.463 | < 0.001 | 1.022 |
| ASJC discipline: HEALTH | 2.039 | 0.010 | 0.937 | 2.177 | 0.029 | 1.367 |
| ASJC discipline: IMMU | −5.787 | −0.037 | 0.720 | −8.040 | 0.000 | 1.044 |
| ASJC discipline: MATER | −4.705 | −0.105 | 0.309 | −15.243 | < 0.001 | 1.535 |
| ASJC discipline: MATH | −4.912 | −0.090 | 0.336 | −14.606 | < 0.001 | 1.356 |
| ASJC discipline: MED | −3.698 | −0.122 | 0.273 | −13.568 | < 0.001 | 2.052 |
| ASJC discipline: PHARM | −5.534 | −0.053 | 0.514 | −10.765 | < 0.001 | 1.094 |
| ASJC discipline: PHYS | −5.663 | −0.105 | 0.338 | −16.744 | < 0.001 | 1.441 |
| ASJC discipline: PSYCH | −1.533 | −0.015 | 0.490 | −3.130 | 0.002 | 1.098 |
| ASJC discipline: SOC | −0.878 | −0.015 | 0.336 | −2.614 | 0.009 | 1.332 |
| ASJC discipline: VET | −4.115 | −0.046 | 0.456 | −9.031 | < 0.001 | 1.119 |
| Academic position: associate professor | 5.407 | 0.234 | 0.114 | 47.567 | < 0.001 | 1.279 |
| Academic position: full professor | 10.923 | 0.357 | 0.171 | 63.708 | < 0.001 | 1.764 |

A few rather interesting findings in the model come from the analysis of the effect of individual academic disciplines on predicted biological age. Scientists and scholars publishing in BUS, DEC, and ENER have similar biological age as those publishing in HUM (which is a reference category). Only assignment to the HEALTH discipline has a significantly positive effect on biological age (by 2 years on average). Publishing in the vast majority of disciplines has a negative effect on biological age compared to HUM (except for the small HEALTH discipline, the small DEC discipline, and the large BUS and ENER disciplines). Assignment to the PHYS, CHEM, IMMU, BIO, PHARM, MATH, MATER, COMP, and MED disciplines has the strongest negative effect on biological age (4–6 years on average). These disciplines belong to the traditional STEMM field, which clearly indicates an earlier start to the academic career (as measured by academic age). The six other STEMM disciplines (i.e., AGRI, ENVIR, CHEMENG, DENT, ENG, EARTH) also have a negative impact compared to HUM by an average of 2–2.5 years. In contrast, the non-STEMM domains (particularly the largest, such as ECON, PSYCH, SOC) also show a negative impact, but clearly smaller than the others (1–1.5 years). This analysis demonstrates the overwhelming supremacy of STEMM disciplines in internationally visible scholarly production.

The analysis of standardized coefficients reveals that the most important predictor of biological age is academic age, the value for which was as high as 0.523 and it was much higher than the second most influential factor—position (associate professor had a value of 0.234, while full professor 0.357). Therefore, these two characteristics are definitely the strongest determinants of biological age. Among the disciplines, CHEM, BIO, MED, PHYS, MATER, and MATH (0.13, 0.13,0.12, 0.11, 0.11, 0.09 respectively) were characterized by relatively high, although much lower than previously mentioned, values of standardized coefficients, which indicates a strong influence of STEMM disciplines on age. The remaining disciplines were characterized by the value of the standardized coefficient significantly below 0.1. Moreover, IDUB did not turn out to be a strong predictor, with a value of only 0.06. The phenomenon of collinearity (significant correlation of the vector of independent variables) did not occur in the model—the values of VIF coefficients in almost every case were lower than 2 (with 4 as a value allowing to state the occurrence of significant interdependence).

In addition, three models were estimated for the age of obtaining PhD, habilitation, and full professorship using the same independent variables. However, these models had a negligible goodness of fit to the empirical data ($R^2$ ranged from 0.005 for the model for the age of PhD to 0.09 for the age of professorship). In fact, the only variable that showed a significant association with each model was academic age, but the magnitude of the coefficient for this variable did not exceed 0.07, thereby implying that an increase in academic age by one year only marginally affects the increase in age at degree completion. In addition, other characteristics such as gender and field significantly affected the age of degree attainment only for the models for habilitation and professorship. These models are not analyzed in this paper.

## Discussion and conclusions

Our detailed examination of the entire population of Polish academic scientists and scholars visible in the last decade in global science and holding at least a PhD ($N = 20,569$) clearly indicates that using academic age as a proxy for biological age in academic career studies works well for STEMM disciplines.

However, for non-STEMM disciplines (particularly for HUM arts and humanities, SOC social sciences, ECON economics, econometrics, and finance, or BUS business, management, and accounting), this usage performs dramatically worse. This negative conclusion is particularly important for systems that are only recently (in the last one or two decades) more widely visible in global academic journals—that is, for countries classified as "developing" (which, apart from Poland, also include Portugal, Slovakia, Bulgaria, Romania, Iran, and Turkey, among others) and as "lagging" (such as Nigeria or Indonesia) (to use Wagner's classification: Wagner, 2008: 88).

Thus, the differences in the usefulness of the usage of academic age in examining academic careers at the micro-level of the individual scientist apply to varying extents to different clusters of countries: on the one hand, there are countries that form the core of global science and on the other hand there are countries that form its peripheries. The idea of core and peripheries (Olechnicka et al., 2019) dates back to the work of Immanuel Wallerstein (1976), but it is increasingly criticized today on the basis of changes in global publication patterns. For example, Marginson (2021) indicates the growing role of newcomers in the global circulation of knowledge, which changes the traditional core-periphery picture and challenges the traditional narratives about where science is produced.

For countries from the scientifically "advanced" cluster, constituting the core of the global circulation of knowledge from the outset, the academic age used as a proxy for biological age works very well for all disciplines (as in Quebec, Canada) (Nane et al. 2017; Costas et al., 2015). In the case of the second cluster of scientifically "developing" countries, represented here by Poland, the proxy works well only for scientists in STEMM disciplines. For STEMM scientists, it works in the 85–90% range—that is, STEMM scientists tend to operate according to globally valid patterns in which the relationship between academic age and biological age is highly predictable. Thus, for example, for scientists working in chemistry, the correlation between academic age and biological age is very high (with correlation coefficient $r = 0.89$), which is similar to physics and astronomy ($r = 0.88$) and mathematics ($r = 0.85$). The correlation is radically lower for non-STEMM scholars for whom the proxy works only in the 35–50% range: it is the lowest for HUM ($r = 0.35$), followed by BUS ($r = 0.42$), ECON ($r = 0.43$), and SOC ($r = 0.49$).

The Polish case confirms the existence of significant differences in global publication patterns in the last three decades between scientifically "developing" countries and "advanced" countries (Wagner, 2008: 88). Poland is a good example of a country that is beyond the traditional academic centers of science and where the number of publications in global academic journals has been systematically increasing. While Western countries have been functioning in common English-language global science networks for several decades, and with radically increasing intensity since 1989, Poland and other scientifically developing countries have been participating in these networks on a larger scale only for the last 10–15 years. Geopolitical differences have resulted in the post-communist countries in the European Union (EU), including Poland (and other poorer countries of the world), becoming massively visible in Scopus or Web of Science journals relatively recently. This is clearly evidenced by the publication data. In contrast, Western countries have been operating on a large scale (and through a high percentage of their scientists and scholars) in the indexed journal system since the very outset. These geopolitical differences have practical implications for the use of academic age in micro-level studies on scientists and their careers.

Our research confirms that the globalization of science is associated with significant cross-disciplinary differentiation between STEMM and non-STEMM disciplines. The individual micro-level data analyzed here suggest a delayed participation of social scientists and humanists in global science networks as opposed to the continuing presence of natural scientists in them. Thus, while science is globalizing fast, it is globalizing unequally: faster in STEMM fields than in non-STEMM fields, with practical implications for predicting biological age from academic age in academic career studies.

Our study supports the idea that it is worthwhile to collect complete data at the individual level: for detailed analyses of academic career in its various dimensions (research productivity, international research collaboration, international mobility), it is useful to have birth dates of all researchers from all science sectors, as in the Polish POL-on system currently under expansion. In such systems, national analyses do not require proxies for biological age—that is, there is no need to use academic age. It is useful to establish and develop full administrative, biographical, and publication databases in the form of Current Research Information Systems (CRIS are institutional and national systems containing data about researchers and research groups, their projects, funding, and research and other outputs) (see Sivertsen, 2019: 667).

Our research also suggests that in scientifically developing countries, academic age as a proxy for biological age must be used more cautiously than in advanced countries and ideally must only be used for STEMM disciplines. The inconsistencies between the two ages in non-STEMM disciplines are greater there and the correlations are radically lower, thereby making the proxy an inadequate analytical tool.

Furthermore, the Polish case refers to several parallel processes that are evident in global science: (1) presence on the peripheries of global science for geopolitical reasons; (2) low but improving individual research productivity; (3) delayed entry into global science and globally indexed academic journals; and (4) permanent underfunding of the academic research system (in the first two decades following 1989). These processes are characteristic of much of the world and affect, to varying degrees, both post-communist countries in Europe and Central Asia, as well as Latin American and African countries.

Academic institutions across the world have complete data on the biological age of their academic staff. In contrast, the availability of such detailed data at the macro level of countries remains limited, thereby impeding the possibility of rigorous analysis of the academic profession in demographic terms. Our study can be replicated for individual institutions, provinces (or other administrative districts), and smaller subpopulations of researchers (such as beneficiaries of grant programs). It can be replicated wherever reliable data on the biological age of researchers and reliable publication data are in place; thus, the results for other institutional, geographic, and national contexts can be compared with ours, thereby leading to a more comprehensive picture. Global comparative studies of academic careers require reliable demographic data on a national scale, the production of which this study strongly encourages.

# Appendix

See Table 5.

**Table 5** Pearson's correlation coefficients for Polish scientists and scholars between the age of being conferred as PhD and academic age and test for association between paired samples statistics in terms of sex, institutional type (research-intensive IDUB institutions and the rest), academic position and ASJC discipline ($N = 20{,}596$)

| Category | Estimate | t-statistic | p-value | Df | Confidence interval—LB | Confidence interval—UB |
|---|---|---|---|---|---|---|
| Total | 0.432 | 67.8 | <0.001 | 20,016 | 0.421 | 0.443 |
| Female scientists | 0.676 | 84.9 | <0.001 | 8556 | 0.664 | 0.687 |
| Male scientists | 0.375 | 43.3 | <0.001 | 11,458 | 0.359 | 0.391 |
| Institution – Rest | 0.371 | 46.8 | <0.001 | 13,722 | 0.357 | 0.386 |
| Institution – IDUB | 0.752 | 90.5 | <0.001 | 6292 | 0.741 | 0.763 |
| Associate Professor | 0.527 | 48.2 | <0.001 | 6059 | 0.508 | 0.545 |
| Full Professor | 0.394 | 22.3 | <0.001 | 2710 | 0.362 | 0.425 |
| Assistant Professor | 0.206 | 22.3 | <0.001 | 11,243 | 0.188 | 0.223 |
| CHEM | 0.900 | 69.2 | <0.001 | 1128 | 0.888 | 0.910 |
| PHYS | 0.888 | 55.1 | <0.001 | 810 | 0.873 | 0.902 |
| IMMU | 0.885 | 17.9 | <0.001 | 89 | 0.830 | 0.923 |
| MATH | 0.867 | 48.7 | <0.001 | 780 | 0.849 | 0.884 |
| PHARM | 0.844 | 22.3 | <0.001 | 201 | 0.799 | 0.879 |
| MATER | 0.843 | 53.4 | <0.001 | 1161 | 0.825 | 0.859 |
| BIO | 0.840 | 58.1 | <0.001 | 1408 | 0.824 | 0.855 |
| DENT | 0.817 | 11.1 | <0.001 | 61 | 0.714 | 0.886 |
| VET | 0.791 | 21.9 | <0.001 | 287 | 0.743 | 0.830 |
| MED | 0.784 | 67.6 | <0.001 | 2873 | 0.769 | 0.797 |
| COMP | 0.773 | 34.4 | <0.001 | 799 | 0.743 | 0.799 |
| EARTH | 0.762 | 35.9 | <0.001 | 931 | 0.734 | 0.788 |
| CHEMENG | 0.751 | 21.8 | <0.001 | 368 | 0.702 | 0.792 |
| DEC | 0.712 | 6.6 | <0.001 | 42 | 0.526 | 0.833 |
| ENVIR | 0.684 | 34.8 | <0.001 | 1380 | 0.655 | 0.711 |
| AGRI | 0.634 | 38.5 | <0.001 | 2199 | 0.609 | 0.659 |
| ENER | 0.590 | 10.9 | <0.001 | 224 | 0.498 | 0.669 |
| PSYCH | 0.522 | 9.3 | <0.001 | 231 | 0.422 | 0.610 |
| SOC | 0.516 | 16.4 | <0.001 | 736 | 0.461 | 0.567 |
| HEALTH | 0.502 | 4.1 | <0.001 | 50 | 0.265 | 0.682 |
| BUS | 0.484 | 12.5 | <0.001 | 513 | 0.415 | 0.548 |
| ECON | 0.416 | 7.6 | <0.001 | 278 | 0.314 | 0.508 |
| HUM | 0.391 | 11.8 | <0.001 | 765 | 0.330 | 0.450 |
| ENG | 0.210 | 11.1 | <0.001 | 2656 | 0.173 | 0.246 |

**Acknowledgements** We gratefully acknowledge the support of the Ministry of Education and Science through its Dialogue Grant 0022/DLG/2019/10 (RESEARCH UNIVERSITIES). We also gratefully acknowledge the assistance of the International Center for the Studies of Research (ICSR) Lab, a cloud-based service provided by Elsevier for research purposes. We are particularly grateful to Kristy James, Data Scientist in the Lab, for her continuous support in this and other projects. We are also grateful to Łukasz Szymula, an excellent doctoral student in the Center for Public Policy Studies. Finally, we want to thank the anonymous reviewers for very constructive criticism.

# References

Abramo, G., Aksnes, D. W., & D'Angelo, C. A. (2020). Comparison of research productivity of Italian and Norwegian professors and universities. *Journal of Informetrics, 14*(2), 101023.

Abramo, G., D'Angelo, C. A., & Solazzi, M. (2011). The relationship between scientists' research performance and the degree of internationalization of their research. *Scientometrics, 86*, 629–643.

Abramo, G., D'Angelo, C. A., & Murgia, G. (2016). The combined effect of age and seniority on research performance of full professors. *Science and Public Policy, 43*(3), 301–319.

Aksnes, D. W., Rørstad, K., Piro, F. N., & Sivertsen, G. (2011a). Age and Scientific Performance. A Large-Scale Study of Norwegian Scientists. In: E., Noyons, P., Ngulube, J., Leta, (Eds.), *Proceedings of ISSI 2011a—the 13th International Conference of the International Society for Scientometrics and Informetrics*, (Vol. 1, pp. 34–45), Durban, South Africa, 4–7 July 2011a, ISSI, Leiden University and University of Zululand, 2011.

Aksnes, D. W., Rørstad, K., Piro, F. N., & Sivertsen, G. (2011b). Are female researchers less cited? A large scale study of Norwegian researchers. *Journal of the American Society for Information Science and Technology, 62*(4), 628–636.

Aksnes, D. W., & Sivertsen, G. (2019). A criteria-based assessment of the coverage of Scopus and web of science. *Journal of Data and Information Science, 4*(1), 1–21. https://doi.org/10.2478/jdis-2019-0001

Antonowicz, D., Kulczycki, E., & Budzanowska, A. (2020). Breaking the deadlock of mistrust? A participative model of the structural reforms in higher education in Poland. *Higher Education Quarterly*. https://doi.org/10.1111/hequ.12254 On-line first February 14, 2020.

Aref, S., Zagheni, E., & West, J., et al. (2019). The demography of the peripatetic researcher: evidence on highly mobile scholars from the web of science. In I. Weber (Ed.), *Social informatics. SocInfo 2019. Lecture Notes in Computer Science.* (Vol. 11864). Berlin: Springer.

Badar, K., Hite, J. M., & Badir, F. Y. (2014). The moderating roles of academic age and institutional sector on the relationship between co-authorship network centrality and academic research performance. *Aslib Journal of Information Management, 66*(5), 38–53. https://doi.org/10.1108/ajim-05-2013-0040

Bieliński, J., & Tomczyńska, A. (2018). The ethos of science in contemporary Poland. *Minerva, 57*(2), 151–173.

Chan, H. F., & Torgler, B. (2020). Gender differences in performance of top cited scientists by field and country. *Scientometrics, 125*, 2421–2447.

Cole, S. (1979). Age and scientific performance. *American Journal of Sociology, 84*(4), 958–977.

Coomes, O. T., Moore, T., Paterson, J., Breau, S., Ross, N. A., & Roulet, N. (2013). Academic performance indicators for departments of geography in the United States and Canada. *The Professional Geographer, 65*(3), 433–450.

Costas, R, Bordons, M. (2007). A classificatory scheme for the analysis of bibliometric profiles at the micro level. *Proceedings of ISSI 2007: 11th international conference of the ISSI, Vols I and II*, pp. 226–230.

Costas, R., Nane, GF., & Lariviere, V. (2015). Is the year of first publication a good proxy of scholars academic age? In A.A. Salah, Y. Tonta, A.A. Akdag Salah (Eds.), *Proceedings of the 15th international conference on scientometrics and informetrics* (pp. 988–998). Istanbul: Bogaziçi University Printhouse.

Costas, R., & Bordons, M. (2005). Bibliometric indicators at the micro-level: Some results in the area of natural resources at the Spanish CSIC. *Research Evaluation, 14*(2), 110–120.

Costas, R., van Leeuwen, T. N., & Bordons, M. (2010a). A bibliometric classificatory approach for the study and assessment of research performance at the individual level: The effects of age on productivity and impact. *Journal of the American Society for Information Science and Technology, 61*(8), 1564–1581.

Costas, R., van Leeuwen, T. N., & Bordons, M. (2010b). Self-citations at the meso and individual levels: Effects of different calculation methods. *Scientometrics, 82*, 517–537.

Elsevier. (2020). *The researcher journey through a gender lens.* Elsevier.

Feldy, M., & Kowalczyk, B. (2020). The ethos of science and the perception of the Polish system of financing science. *European Review, 28*(4), 599–616.

Gingras, Y., Larivière, V., Macaluso, B., & Robitaille, J. P. (2008). The effects of aging on researchers' publication and citation patterns. *PLoS ONE, 3*(12), e4048. https://doi.org/10.1371/journal.pone.0004048 Epub 2008 Dec 29. PMID: 19112502; PMCID: PMC2603321.

Guns, R., Eykens, J., & Engels, T. C. E. (2019). To what extent do successive cohorts adopt different publication patterns? Peer review, language use, and publication types in the social sciences and humanities. *Frontiers in Research Metrics and Analytics.* https://doi.org/10.3389/frma.2018.0003

GUS. (2020). *Higher education institutions and their finances in 2019.* GUS. Central Statistical Office.

Harzing, A. W. (2019). Two new kids on the block: How do Crossref and Dimensions compare with Google Scholar, Microsoft Academic, Scopus and the Web of Science? *Scientometrics, 120*(1), 341–349.

Kulczycki, E., Guns, R., Pölönen, J., Engels, T. C. E., Rozkosz, E. A., Zuccala, A. A., & Sivertsen, G. (2020). Multilingual publishing in the social sciences and humanities: A seven-country European Study. *Journal of the Association for Information Science and Technology., 71*(11), 1371–1385. https://doi.org/10.1002/asi.24336

Kulczycki, E., Korzeń, M., & Korytkowski, P. (2017). Toward an excellence-based research funding system: Evidence from Poland. *Journal of Informetrics, 11*(1), 282–298.

Kwiek, M. (2015a). The internationalization of research in Europe. A quantitative study of 11 national systems from a micro-level perspective. *Journal of Studies in International Education, 19*(2), 341–359. https://doi.org/10.1177/1028315315572898

Kwiek, M. (2015b). Academic generations and academic work: Patterns of attitudes, behaviors and research productivity of Polish academics after 1989. *Studies in Higher Education, 40*(8), 1354–1376. https://doi.org/10.1080/03075079.2015.1060706

Kwiek, M. (2018a). High research productivity in vertically undifferentiated higher education systems: Who are the top performers? *Scientometrics, 115*(1), 415–462. https://doi.org/10.1007/s11192-018-2644-7

Kwiek, M. (2018b). International research collaboration and international research orientation: Comparative findings about European academics. *Journal of Studies in International Education, 22*(1), 1–25. https://doi.org/10.1177/1028315317747084

Kwiek, M. (2019). *Changing European academics. A comparative study of social stratification, work patterns and research productivity.* Routledge.

Kwiek, M. (2020). Internationalists and locals: International research collaboration in a resource-poor system. *Scientometrics, 124*, 57–105. https://doi.org/10.1007/s11192-020-03460-2

Kwiek, M. (2021). What large-scale publication and citation data tell us about international research collaboration in Europe: Changing national patterns in global contexts. *Studies in Higher Education, 46*(12), 2629–2649. https://doi.org/10.1080/03075079.2020.1749254

Kwiek, M., & Roszka, W. (2021a). Gender disparities in international research collaboration: A large-scale bibliometric study of 25,000 university professors. *Journal of Economic Surveys, 35*(5), 1344–1388. https://doi.org/10.1111/joes.12395

Kwiek, M., & Roszka, W. (2021b). Gender-based homophily in research: A large-scale study of man-woman collaboration. *Journal of Informetrics, 15*(3), 1–38. https://doi.org/10.1016/j.joi.2021.101171

Kwiek, M., & Roszka, W. (2022). Are female scientists less inclined to publish alone? The gender solo research gap. *Scientometrics.* https://doi.org/10.1007/s11192-022-04308-7 Online first, March 08, 2022.

Kwiek, M., & Szadkowski, K. (2019). Higher education systems and institutions: Poland. In P. N. Texteira & J. C. Shin (Eds.), *International encyclopedia of higher education systems* (pp. 1–20). Springer.

Kyvik, S. (1990). Age and scientific productivity. Differences between fields of learning. *Higher Education, 19*, 37–55.

Kyvik, S., & Olsen, T. B. (2008). Does the aging of tenured academic staff affect the research performance of universities? *Scientometrics, 76*(3), 439–455.

Larivière, V., Vignola-Gagné, E., Villeneuve, C., et al. (2011). Sex differences in research funding, productivity and impact: An analysis of Québec university professors. *Scientometrics, 87*, 483–498.

Lee, S. Bozeman, B. (2005). The Impact of Research Collaboration on Scientific Productivity. *Social Studies of Science, 35*(5), 673–702.

Lehman, H. C. (1953). *Age and achievement*. Princeton University Press.

Levin, S., Stephan, P.E. (1991). Research productivity over the life cycle: Evidence for academic scientists. *The American Economic Review, March 1991*, 114–132.

Liao C.H. (2017). Reopening the black box of career age and research performance. In J. Zhou, G. Salvendy, (Eds), *Human Aspects of IT for the Aged Population. Applications, Services and Contexts. ITAP 2017. Lecture Notes in Computer Science* (vol 10298). Springer. https://doi.org/10.1007/978-3-319-58536-9_41

Marginson, S. (2021). What drives global science? The four competing narratives. *Studies in Higher Education*. https://doi.org/10.1080/03075079.2021.1942822

Milojević, S. (2012). How are academic age, productivity, and collaboration related to citing behavior of researchers? *PLoS ONE, 7*(11), e49176. https://doi.org/10.1371/journal.pone.0049176 Epub 2012 Nov 7. PMID: 23145111; PMCID: PMC3492318.

Mongeon, P., & Paul-Hus, A. (2016). The journal coverage of Web of Science and Scopus: A comparative analysis. *Scientometrics, 106*, 213–228. https://doi.org/10.1007/s11192-015-1765-5

Nane, G. F., Larivière, V., & Costas, R. (2017). Predicting the age of researchers using bibliometric data. *Journal of Informetrics, 11*(3), 713–729.

Olechnicka, A., Ploszaj, A., & Celinska-Janowicz, D. (2019). *The geography of scientific collaboration*. Routledge.

Pelz, D. C., & Andrews, F. W. (1976). *Scientists in organizations*. Wiley.

Perianes-Rodriguez, A., & Ruiz-Castillo, J. (2014). Within- and between-department variability in individual productivity: The case of economics. *Scientometrics, 102*(2), 1497–1520.

Petersen, A. M. (2015). On the impact of super ties in scientific careers. *Proceedings of the National Academy of Sciences, 112*(34), E4671–E4680. https://doi.org/10.1073/pnas.1501444112

Petr, M., Engels, T. C. E., Kulczycki, E., Dušková, M., Guns, R., Sieberová, M., et al. (2021). Journal article publishing in the social sciences and humanities: A comparison of Web of Science coverage for five European countries. *PLoS One, 16*(4), e0249879. https://doi.org/10.1371/journal.pone.0249879

Radicchi, F., & Castellano, C. (2013). Analysis of bibliometric indicators for individual scholars in a large data set. *Scientometrics, 97*, 627–637.

Robinson-Garcia, N., Costas, R., Sugimoto, C. R., Larivière, V., & Nane, G. F. (2020). Task specialization across research careers. *eLife, 9*, e60586. https://doi.org/10.7554/eLife.60586

Rørstad, K., & Aksnes, D. W. (2015). Publication rate expressed by age, gender and academic position—A large-scale analysis of Norwegian academic staff. *Journal of Informetrics, 9*, 317–333.

Rørstad, K., Aksnes, D. W., & Piro, F. N. (2021). Generational differences in international research collaboration: A bibliometric study of Norwegian University staff. *PLoS ONE, 16*(11), e0260239. https://doi.org/10.1371/journal.pone.0260239

Savage, W. E., & Olejniczak, A. J. (2021). Do senior faculty members produce fewer research publications than their younger colleagues? Evidence from Ph.D. granting institutions in the United States. *Scientometrics, 126*, 4659–4686. https://doi.org/10.1007/s11192-021-03957-4

SciVal (2021). Global bibliometric dataset. https://www.scival.com (restricted access)

Shaw, M. A. (2019). Strategic instrument or social institution: Rationalized myths of the university in stakeholder perceptions of higher education reform in Poland. *International Journal of Educational Development, 69*, 9–21.

Simoes, N., & Crespo, N. (2020). A flexible approach for measuring author-level publishing performance. *Scientometrics, 122*, 331–355.

Singh, V. K., Singh, P., Karmakar, M., Leta, J., & Mayr, P. (2021). The journal coverage of Web of Science, Scopus and Dimensions: A comparative analysis. *Scientometrics, 126*(6), 5113–5142. https://doi.org/10.1007/s11192-021-03948-5

Sivertsen, G. (2019). Developing current research information systems (CRIS) as data sources for studies of research. In W. Glänzel, H. F. Moed, U. Schmoch, & M. Thelwall (Eds.), *Springer handbook of science and technology indicators*. Springer.

Stephan, P. (2012). *How economics shapes science*. Harvard University Press.

Stephan, P. E., & Levin, S. G. (1992). *Striking the mother lode in science: The importance of age, place, and time*. Oxford University Press.

Stern, N. (1978). Age and achievement in mathematics: A case-study in the sociology of science. *Social Studies of Science, 8*(1), 127–140.

Sugimoto, C. R., Sugimoto, T. J., Tsou, A., Milojević, S., & Larivière, V. (2016). Age stratification and cohort effects in scholarly communication: A study of social sciences. *Scientometrics, 109*(2), 997–1016. https://doi.org/10.1007/s11192-016-2087-y

Van den Besselaar, P., & Sandström, U. (2016). Gender differences in research performance and its impact on careers: A longitudinal case study. *Scientometrics, 106*(1), 143–162.

Wagner, C. S. (2008). *The new invisible college*. Brookings Institution Press.

Wais, K. (2016). Gender Prediction Methods Based on First Names with genderizeR. *The R Journal, 8*(1), 17–37.

Wallerstein, I. (1976). Semi-peripheral countries and the contemporary world crisis. *Theory and Society, 3*, 461–483.

Wang, D., & Barabási, A. (2021). *The science of science*. Cambridge University Press. https://doi.org/10.1017/9781108610834

Wildgaard, L. (2015). A comparison of 17 author-level bibliometric indicators for researchers in Astronomy, Environmental Science, Philosophy and Public Health in Web of Science and Google Scholar. *Scientometrics, 104*, 873–906.

Zuckerman, H., & Merton, R. K. (1973). Age, aging, and age structure in science. In M. W. Riley, M. Johnson, & A. Foner (Eds.), *Aging and society, vol. 3. A sociology of age stratification.* Russell Sage Foundation.